\documentclass[11pt,technote,onecolumn]{IEEEtran}
\usepackage{cite,amsmath,graphicx}
\usepackage{bm}
\usepackage{amssymb}
\usepackage[ruled]{algorithm2e}
\usepackage{booktabs}
\usepackage{multirow}
\usepackage{epstopdf}
\usepackage{amsthm}
\usepackage{lineno}
\usepackage{mathrsfs}
\usepackage{amsmath}
\usepackage{txfonts}
\usepackage{graphicx}
\usepackage{caption}
\usepackage{url}

\usepackage{amsmath,graphicx}

\usepackage{amsfonts}

\usepackage[justification=centering]{caption}
\usepackage[ruled]{algorithm2e}
\usepackage{subfigure}
\usepackage{verbatim}

\usepackage{subfigure}
\ifCLASSINFOpdf
\else
\fi
\begin{document}
%
\title{Hyperspectral Image Denoising with Partially Orthogonal Matrix Vector Tensor Factorization}
%
%
%

\author{Zhen Long,
        Yipeng Liu,~\IEEEmembership{Member,~IEEE,}
       Sixing Zeng, Jiani Liu, Fei Wen, Ce Zhu, ~\IEEEmembership{Fellow,~IEEE,} 

\thanks{This research is supported by National Natural Science Foundation of China (NSFC, No. 61602091, No. 61571102) and the Sichuan Science and Technology program  (No. 2019YFH0008, No. 2018JY0035). The corresponding author is Yipeng Liu.}

\thanks{Z. Long, Y. Liu, S. Zeng, J. Liu and C. Zhu are with School
of Communication and Information Engineering, University of Electronic Science and Technology of China, Chengdu, 611731, China.  e-mail: yipengliu@uestc.edu.cn.}
\thanks{F. Wen is with the Department of Electronic Engineering, Shanghai
	Jiao Tong University, Shanghai 200240, China.}
}

%
%

\markboth{Journal of \LaTeX\ Class Files,~Vol.~xx, No.~xx, month~year}%
{Shell \MakeLowercase{\textit{et al.}}: Bare Demo of IEEEtran.cls for IEEE Journals}
%



\maketitle

\begin{abstract}
Hyperspectral image (HSI) has some advantages over  natural image for various applications due to the extra spectral information. During the acquisition, it is often contaminated by severe noises including Gaussian noise, impulse noise, deadlines, and stripes. The image quality degeneration would badly effect some applications. In this paper, we present an HSI restoration method named smooth and robust low-rank tensor recovery. Specifically, we propose a structural tensor decomposition in accordance with the linear spectral mixture model of HSI. It decomposes a tensor into sums of outer matrix-vector products, where the vectors are orthogonal due to the independence of endmember spectrums. Based on it, the global low rank tensor structure can be well exposited for HSI denoising.  In addition,  the 3D anisotropic total variation is used for spatial-spectral piecewise smoothness of HSI. Meanwhile, the sparse noise including impulse noise, deadlines and stripes, is detected by the $\ell_1$-norm regularization. The Frobenius norm is used for the heavy Gaussian noise in some real-world scenarios. The alternating direction method of multipliers is adopted to solve the proposed optimization model, which simultaneously exploits the global low-rank property and the spatial–spectral smoothness of the HSI. Numerical experiments on both simulated and real data  illustrate the superiority of the proposed method in comparison with the existing ones.

\end{abstract}


%
\IEEEpeerreviewmaketitle

\section{Introduction}
%
%
%
%
\IEEEPARstart{H}{yperspectral} imaging  uses spectrometers to collect data over hundreds of spectral bands  ranging from 400nm to 2500nm in the same region.  It can provide both spectral and spatial information about objects due to its numerous and continuous spectral bands. With abundant available spectral information, hyperspectral image (HSI) has been popular in a series of fields, such as remote sensing~\cite{jensen1987introductory}, food safety~\cite{gowen2007hyperspectral}, object detection and classification~\cite{tarabalka2010segmentation, manolakis2002detection}. However, due to thermal electronics, dark current, random error of light count  in the image formation, HSI is inevitably affected by severe noise during the acquisition, and it will definitely degrade the image quality. Therefore, the noise removal from HSI has become an important research area and attracted lots of attentions~\cite{yuan2012hyperspectral, zhao2014hyperspectral, rasti2012hyperspectral, he2015hyperspectral, yuan2013hyperspectral}.

\begin{figure}[h]
\centering
\includegraphics[scale=0.11]{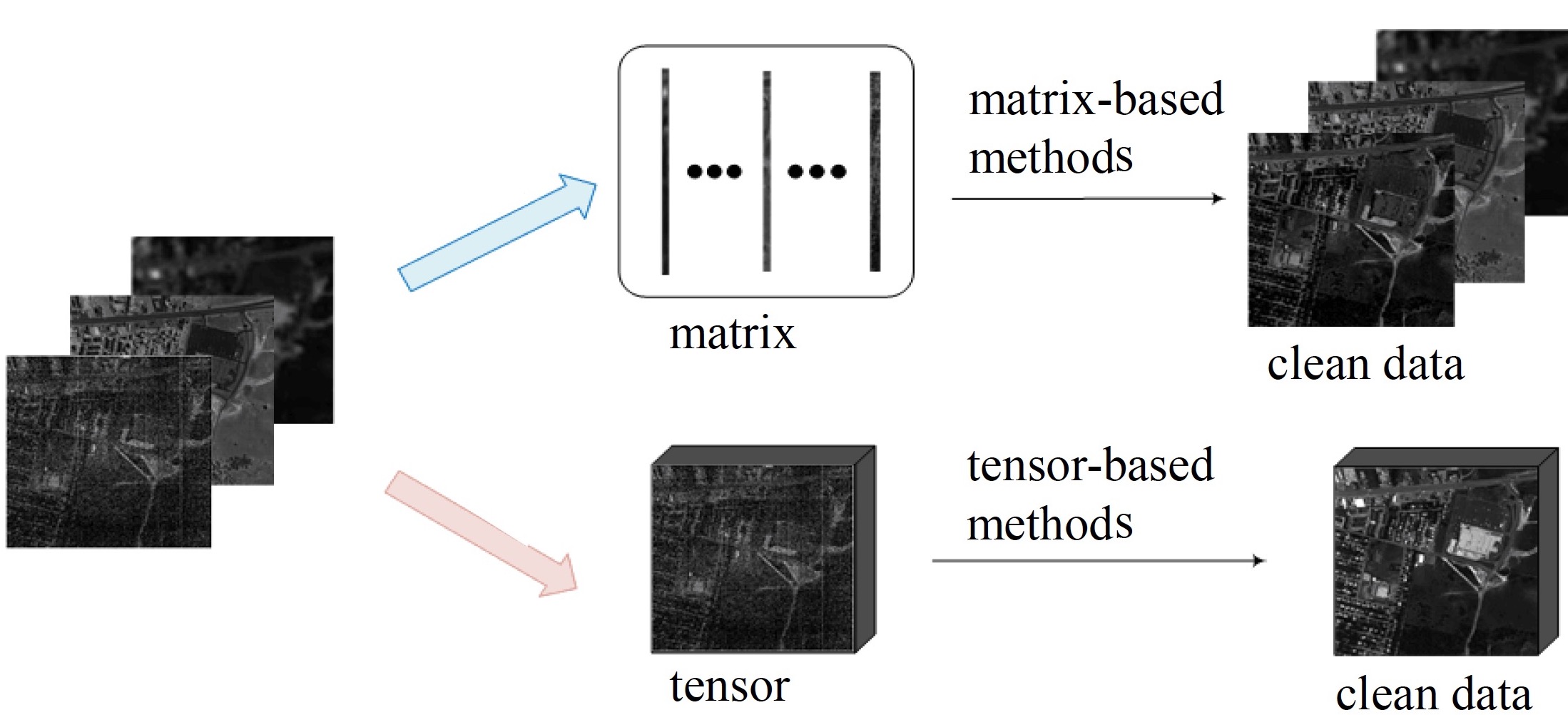}
\caption{The schematic diagram of matrix based methods and tensor  based methods}
\label{1}
\end{figure}

A number of  denoising algorithms have proposed for HSI, including block matching and 3-dimensional filtering (BM3D) \cite{dabov2006image}, total variational based methods \cite{bresson2008fast,ijgi7100412,aggarwal2016hyperspectral},  sparse representations based methods \cite{ elad2006image,yan2010visual, tian2017cross},  Bayesian methods \cite{ lebrun2013nonlocal,xu2015hyperspectral,zheng2018student, gu2017structural}), deep learning  based methods \cite{xie2017hyperspectral, wang2016temperature,qu2016multilevel,yuan2018hyperspectral,maffei2019single}, and low rank based methods~\cite{zhang2013hyperspectral,zhu2014spectral,he2015hyperspectral,he2018non,fan2017hyperspectral}. Among them, low rank approximation shows good performance without training, as a representative of model-based ways. It can be shown in Fig. \ref{1},  low-rank based denoising methods can be divided into matrix based ones and tensor based ones.

The matrix-based denoising unfolds the three-dimensional tensor into a matrix or treats each band independently, and uses the traditional two-dimensional image denoising methods. For example, Zhang et al.~\cite{zhang2013hyperspectral} divide the HSI into several fragments, rearrange these fragments into a two-dimensional matrix, and restore each fragment using the low rank matrix recovery (LRMR) method. This method is very successful in recovering HSI with mixed noise, leading to a series of denoising models based on LRMR \cite{he2015hyperspectral, zhu2014spectral, he2018non}.

However, these two-dimensional denoising algorithms are difficult to achieve optimal results since the joint spatial-spectral information of HSI is partly damaged. To address this problem, the low-rank tensor recovery (LRTR)~\cite{huang2020provable}  is proposed for HSI denoising~\cite{fan2017hyperspectral}, which simultaneously makes use of the spectral and spatial information of HSI and obtains better results than the LRMR methods. Following it, some other works using different tensor decompositions are proposed. For instance, LRTDTV \cite{8233403} uses Tucker decomposition based low rank tensor recovery method~\cite{kim2007nonnegative,long2019low} for HSI denoising. NLR-CPTD \cite{xue2019nonlocal} and GSLRTD \cite{huang2018hyperspectral}  apply CP decomposition~\cite{jiang2004kruskal} and  t-SVD~\cite{kilmer2013third,feng2020robust} for HSI denoising, respectively.

\begin{figure}[htbp]
\begin{center}
\includegraphics[scale=0.2]{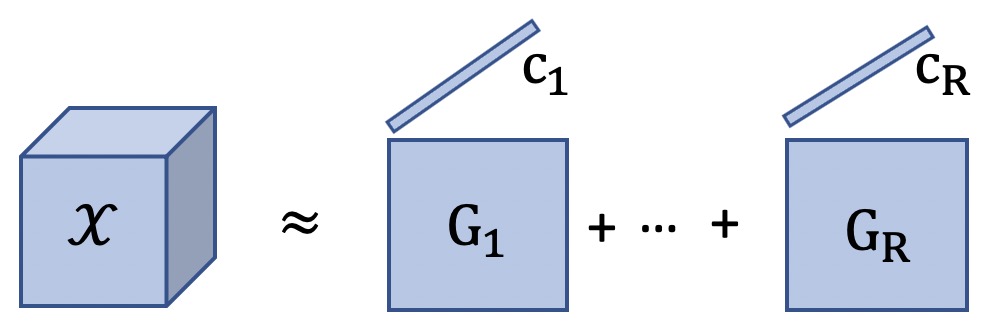}
\caption{Matrix-vector tensor factorization}
\label{default}
\end{center}
\end{figure}

However, the commonly used three tensor decompositions cannot explicitly represent the linear spectral mixture model for HSI processing. In linear spectral mixture model, a spectrum at each pixel  in the spatial domain is assumed to be a linear combination of several endmember spectra. Matrix-vector tensor factorization (MVTF) provides a more natural way to fit the linear spectral mixture model of HSI. As shown in Fig. 2, a HSI can be represented by the sum of the cross products of a endmember (vector) and its abundance map (matrix). Besides, MVTF has shown its superiority on HSI related applications, such as HSI unmixing~\cite{7784711} and HSI super-resolution~\cite{zhang2019hyperspectral}.  
However, in these two cases, they all ignore that the independence of endmember spectrums which is an important linear spectral mixture model prior for HSI related applications~\cite{wang2014abundance,nascimento2005does}.	


In this paper, motivated by the linear spectral mixture model, we propose a structural MVTF with orthogonal vectors, and apply it to low rank based HSI denoising. The newly obtained partially orthogonal  matrix vector tensor factorization can better exploit the global structure of HSI. To further enhance the recovery performance, total variation (TV)~\cite{liu2020smooth,liu2018image} is used to model the piece-smooth structure of HSI. In addition, the Frobenius norm and $\ell_1$ norm terms are used to deal with Gaussian noise and impulse noise in the optimization model, respectively.
The alternating direction method of multipliers (ADMM) is used to divide the optimization problem into several sub-problems. Experimental results based on HSI recovery show that the proposed method outperforms state-of-the-art ones in terms of mean peak signal to noise ratio (MPSNR), mean structural similarity
index (MSSIM), and dimensionless global relative error of synthesis (ERGAS).

The main contributions of this paper can be summarized as follows:
\begin{itemize}
\item[1)] We develop a new low rank approximation method based on a new structural tensor factorization to model HSI. Considering that endmember spectrums are irrelevant, an orthogonality constraint is introduced on vectors (endmember spectrums) in MVTF. With this constraint, we directly minimize the ranks of abundance matrices. Compared with the method on standard MVTF, the proposed one is much faster and there is no need to set all ranks of MVTF in advance.

\item[2)]A designed 3D anisotropic total variation (3DATV) term for spatial-spectral regularization is incorporated into the optimization model for low-rank tensor recovery. It can well model the local piece-wise smoothness in joint spatial-spectral domain of HSI data for denoising.

\item[3)] The ADMM is used for solving the optimization model. Numerical experiments on simulated and real data illustrate the superiority of the proposed methods.
\end{itemize}




	The structure of the paper is as follows. In Section 2, we give some notations used in this paper. Section 3 presents the details of the proposed hyperspectral image denoising method. The experimental results of the proposed method is given in Section 4. Section 5 summarizes the paper.

\section{Notations and Preliminaries}

\subsection{Notations}
In this paper, scalar, vector, matrix and tensor are denoted by lowercase letters $x$, boldface lowercase letters $\mathbf{x}$ and boldface capital letters $\mathbf{X}$, and  calligraphic letters $\mathcal{X}$, respectively.


\subsection{Preliminaries on tensor computation}

The $\ell_1$ norm and the Frobenius norm of a matrix $\mathbf{X}$ are defined as:

\begin{equation}
\begin{split}
\left \| \mathbf{X} \right \|_{1}=\sum^{I_1}_{i_1=1}\sum ^{I_2}_{i_2=1}\left | x_{i_1,i_2} \right |,
\end{split}
\end{equation}
and
\begin{equation}
\begin{split}
\left \| \mathbf{X} \right \|_{\text{F}}=(\sum^{I_1}_{i_1=1}\sum ^{I_2}_{i_2=1}\left | x_{i_1,i_2} \right |^2)^{1/2},
\end{split}
\end{equation}
respectively, where $x_{i_1,i_2}$ is the $(i_1,i_2)$-th entry of $\mathbf{X}$.

The inner product of two tensors $\mathcal{X},\mathcal{Y}\in\mathbb{R}^{I_1\times I_2\times \cdots\times I_N}$ is defined as:
\begin{equation}
\begin{split}
\left \langle \mathcal{X},\mathcal{Y} \right \rangle=\sum ^{I_1}_{i_1=1}\sum ^{I_2}_{i_2=1}\cdots\sum ^{I_N}_{i_N=1}x_{i_1\cdots i_N}y_{i_1\cdots i_N}.
\end{split}
\end{equation}

The cross product of two tensors $\mathcal{X}\in\mathbb{R}^{I_1\times I_2\times \cdots\times I_N}$ and  $\mathcal{Y}\in\mathbb{R}^{J_1\times J_2\times \cdots\times J_M}$ is defined as:
\begin{equation}
(\mathcal{X}\circ\mathcal{Y})_{i_1\cdots i_Nj_1\cdots j_M}=x_{i_1\cdots i_N}y_{j_1\cdots j_M} .
\end{equation}

The mode-$n$ product of an $N$-th order tensor $\mathcal{X}\in\mathbb{R}^{I_1\times I_2\times\cdots\times I_N}$ with a matrix $\mathbf{U}\in\mathbb{R}^{J_n\times I_n}$ is a new tensor of size $I_1\times\cdots\times I_{n-1}\times J_n\times I_{n+1}\times\cdots\times I_N$, which can be denoted as follows:
\begin{equation}
\begin{split}
(\mathcal{X}\times_n\mathbf{U}^{(n)})_{i_1\cdots i_{n-1}ji_{n+1}\cdots i_N}=\sum _{i_n=1}^{I_n}x_{i_1 i_2\cdots i_N}u_{ji_n},
\end{split}
\end{equation}

The mode-$n$ unfolding of an $N$-th order tensor $\mathcal{X}\in\mathbb{R}^{I_1\times I_2\times\cdots\times I_N}$  is expressed as $\mathbf{A}_{(n)}$. The mode-$n$ unfolding operator arranges the $n$-th mode of $\mathcal{A}$ as the row while the rest modes as the column of the mode-$n$ unfolding matrix. Mathematically, the elements of $\mathbf{A}_{(n)}$ satisfy
\begin{equation}
\mathbf{A}_{(n)}(i_{n},j)=\mathcal{A}(i_{1}, \cdots, i_{n}, \cdots, i_{N})
\end{equation}
where $j=\overline{i_{1},\cdots,i_{n-1},i_{n+1},\cdots,i_{N}}$.

\subsection{Preliminaries on  total variation}

%
For a third-order hyperspectral data $\mathcal{X}\in\mathbb{R}^{I\times J\times K}$, the total variation of HSI is denoted by~\cite{  fan2018spatial, wang2017hyperspectral}.:
\begin{eqnarray}\label{TV}
&&\|\mathcal{X}\|_{\text{TV}}=\|\mathfrak{D}(\mathcal{X})\|_{1}\nonumber\\
&&=\|\mathfrak{D}_{h}(\mathcal{X})\|_{1}+\|\mathfrak{D}_{v}(\mathcal{X})\|_{1}+\|\mathfrak{D}_{z}(\mathcal{X})\|_{1},
\end{eqnarray}
where $\mathfrak{D}(\cdot)=\left [\mathfrak{D}_{h}(\cdot);\mathfrak{D}_{v}(\cdot);\mathfrak{D}_{z}(\cdot) \right ]$ is a three-dimensional difference operator, $\mathfrak{D}_{h}$, $\mathfrak{D}_{v}$, $\mathfrak{D}_{z}$ are first-order finite difference operators along three different ways of hyperspectral data, which can be defined as:
\begin{eqnarray}
&&\mathfrak{D}_{h}(\mathcal{X})=\mathcal{X}(i+1,j,k)-\mathcal{X}(i,j,k), i=1,\cdots,I-1\nonumber\\
&&\mathfrak{D}_{v}(\mathcal{X})=\mathcal{X}(i,j+1,k)-\mathcal{X}(i,j,k), j=1,\cdots,J-1\nonumber\\
&&\mathfrak{D}_{z}(\mathcal{X})=\mathcal{X}(i,j,k+1)-\mathcal{X}(i,j,k), k=1,\cdots,K-1.
\end{eqnarray}




\subsection{ Matrix-vector tensor factorization}
The  MVTF decomposes a tensor $\mathcal{X}\in\mathbb{R}^{I\times J\times K}$ into the sum of several component tensors, and each component tensor can be written in the cross product form of a matrix $\mathbf{G}_r\in \mathbb{R}^{I\times J}$ and a vector $\mathbf{c}_r\in\mathbb{R}^{K}$, which can be denoted as~\cite{7784711}:
\begin{equation}
\mathcal{X}\approx \sum _{r=1}^{R}\mathbf{G}_r\circ \mathbf{c}_r=\mathcal{G}\times_{3}\mathbf{C},
\end{equation}
where $\mathcal{G}\in \mathbb{R}^{I\times J\times R}$, $\mathbf{G}_r$ is the $r$-th slice of $\mathcal{G}$ and $\mathbf{C}\in \mathbb{R}^{K\times R}$.

In linear spectral mixture model of hyperspectral image, $\mathbf{c}_r$ can be modeled as the  $r$-th endmember spectrum, and abundance matrix $\mathbf{G}_r$ represents the spatial information of $r$-th endmember.  Due to the strong spatial correlation of HSI, each abundance matrix is always low rank.  

\section{Optimization model}
As mentioned above, HSI is a third-order tensor, which has two spatial dimensions and one spectral dimension.  In the data acquisition, they are inevitably polluted by noise.
Considering that the observed image $\mathcal{Y}$ can be polluted by Gaussian noise and outliers, we set up the measurement model as follows:
\begin{equation}
\mathcal{Y}=\mathcal{X}+\mathcal{S}+\mathcal{N},
\end{equation}
where $\mathcal{Y}$ is the observed image, $\mathcal{X}$ is the  clean image, $\mathcal{S}$ is the sparse noise or outliers, and $\mathcal{N}$ is the Gaussian noise.

The purpose of HSI denoising is to estimate the clear hyperspectral image from the  observed  data contaminated by noise. In this study, we propose a novel smooth and robust low rank partially orthogonal MVTF method  for HSI denoising under the consideration that endmember spectrums are irrelevant.

 First, we develop a low rank partially orthogonal MVTF model for the clean HSI. It can be achieved by the following model:
\begin{equation}
\min_{\mathbf{G}_{r,r=1,\cdots R}, \mathbf{C}}\sum_{r=1}^{R}\operatorname{rank}(\mathbf{G}_r),~\text{s.~t.~} \mathcal{X}=\mathcal{G}\times_3\mathbf{C},\mathbf{C}^{\operatorname{T}}\mathbf{C}=\mathbf{I}_R,
\end{equation}
where $\mathcal{X}$ is the clean HSI, $\operatorname{rank}(\mathbf{G}_r)$ is the rank of $r$-th abundance matrix $\mathbf{G}_r$, $\mathcal{G}(:,:,r)=\mathbf{G}_r$,  $\min_{\mathbf{G}_{r,r=1,\cdots R},\mathbf{C}}\sum_{r=1}^{R}\operatorname{rank}(\mathbf{G}_r)$ means each abundance matrix is of low rank, and the term $\mathbf{C}^{\operatorname{T}}\mathbf{C}=\mathbf{I}_R$ means endmember spectrums are uncorrelated. Next,  the $ \ell_1 $ norm is used to separate sparse outlier from the observation. In addition, to enhance the recovery performance, the total variation term is  used to exploit local smoothness structure of the HSI data.

The optimization model for this smooth and robust low rank tensor recovery (SRLRTR) for HSI denoising can be formulated as follows:
\begin{eqnarray}\label{equation:mvtf}
&&\min_{\mathcal{X}, \mathcal{S}, \mathcal{N}, \mathcal{G}, \mathbf{C}}\lambda_{\text{TV}}\left \| \mathcal{X} \right \|_{\text{TV}}+\lambda_{\mathcal{S}}\left \| \mathcal{S} \right \|_{1}+\lambda_{\mathcal{N}}\left \| \mathcal{N} \right \|_{\text{F}}^{2}+\lambda_{\mathcal{G}}\sum_{r=1}^{R}\operatorname{rank}(\mathbf{G}_r)\nonumber\\
&& \text{s.~t.} \quad \mathcal{Y}=\mathcal{X}+\mathcal{S}+\mathcal{N}, \mathcal{X}=\mathcal{G}\times_3\mathbf{C},\mathbf{C}^{\operatorname{T}}\mathbf{C}=\mathbf{I}_R.
\end{eqnarray}

Since the function $\operatorname{rank}(\mathbf{G}_r)$ is nonconvex, we can use $\|\mathbf{G}_r\|_{*}$ as the convex surrogate and rewrite equation (\ref{equation:mvtf}) as follows:
\begin{eqnarray}\label{equation:mvtf1}
&&\min_{\mathcal{X},\mathcal{S},\mathcal{N},\mathcal{G},\mathbf{C}}\lambda_{\text{TV}}\left \| \mathfrak{D}(\mathcal{X}) \right \|_{1}+\lambda_{\mathcal{S}}\left \| \mathcal{S} \right \|_{1}+\lambda_{\mathcal{N}}\left \| \mathcal{N} \right \|_{\text{F}}^{2}+\lambda_{\mathcal{G}}\sum_{r=1}^{R}\|\mathbf{G}_r\|_*\nonumber\\
&& \text{s.~t.} \quad \mathcal{Y}=\mathcal{X}+\mathcal{S}+\mathcal{N}, \mathcal{X}=\mathcal{G}\times_3\mathbf{C},\mathbf{C}^{\operatorname{T}}\mathbf{C}=\mathbf{I}_R.
\end{eqnarray}

For solving this problem, we need to introduce two auxiliary variables $\mathcal{Z}$ and $\mathbf{L}$, and the optimization model (\ref{equation:mvtf1}) can be rewritten into the following equivalent form:
\begin{eqnarray}\label{equation:mvtf2}
&&\min_{\mathcal{X},\mathcal{S},\mathcal{N},\mathcal{G},\mathbf{C},\mathcal{Z},\mathbf{L}}\lambda_{\text{TV}}\left \| \mathbf{L}\right \|_{1}+\lambda_{\mathcal{S}}\left \| \mathcal{S} \right \|_{1}+\lambda_{\mathcal{N}}\left \| \mathcal{N} \right \|_{\text{F}}^{2}+\lambda_{\mathcal{G}}\sum_{r=1}^{R}\|\mathbf{G}_r\|_*\nonumber\\
&& \text{s.~t.} \quad \mathcal{Y}=\mathcal{X}+\mathcal{S}+\mathcal{N}, \mathcal{Z}=\mathcal{X},\mathbf{L}=\mathfrak{D}(\mathcal{Z}) ,\nonumber\\
&&\quad\mathcal{X}=\mathcal{G}\times_3\mathbf{C}, \mathbf{C}^{\operatorname{T}}\mathbf{C}=\mathbf{I}_R.
\end{eqnarray}
The  augmented Lagrangian function of the optimization model (\ref{equation:mvtf2}) is:
\begin{eqnarray}\label{equation:mvtf3}
&&\mathfrak{L}(\mathcal{X},\mathcal{S},\mathcal{N},\mathcal{G},\mathbf{C},\mathcal{Z},\mathbf{L},\Lambda_1,\Lambda_2,\Lambda_3,\Lambda_4)=\lambda_{\text{TV}}\left \| \mathbf{L}\right \|_{1}+\lambda_{\mathcal{S}}\left \| \mathcal{S} \right \|_{1}\nonumber\\
&&+\lambda_{\mathcal{N}}\left \| \mathcal{N} \right \|_{\text{F}}^{2}+\lambda_{\mathcal{G}}\sum_{r=1}^{R}\|\mathbf{G}_r\|_*+\left \langle \Lambda _1,\mathcal{Y}-\mathcal{X} -\mathcal{S}-\mathcal{N}\right \rangle\nonumber\\
&&+\left \langle \Lambda _2,\mathcal{Z}-\mathcal{X} \right \rangle+\left \langle \Lambda _3,\mathbf{L} -\mathfrak{D}(\mathcal{Z})\right \rangle+\left \langle \Lambda _4,\mathcal{X}-\mathcal{G}\times_3\mathbf{C} \right \rangle\nonumber\\
&&+\frac{\beta_{1}}{2}\left \| \mathcal{Y}-\mathcal{X}-\mathcal{S}-\mathcal{N} \right \|_\text{F}^2+\frac{\beta_2}{2}\left \| \mathcal{Z}-\mathcal{X} \right \|_\text{F}^2\nonumber\\
&&+\frac{\beta_3}{2}\left \| \mathbf{L} -\mathfrak{D}(\mathcal{Z})\right \|_\text{F}^2+\frac{\beta_4}{2}\left \| \mathcal{X}-\mathcal{G}\times_3\mathbf{C} \right \|_\text{F}^2),
\end{eqnarray}
under the constraint $\mathbf{C}^{\operatorname{T}}\mathbf{C}=\mathbf{I}_R$. Therefore, we can choose to iteratively optimize  one variable with the other variables fixed.

\subsection{Updating $\mathcal{G}$}
Fixing other variables to update $\mathcal{G}$, the optimization model can be rewritten  as:
\begin{equation}\label{equation:update G}
\min_{\mathbf{G}_{r,r=1,\cdots,R}}  \lambda_{\mathcal{G}}\sum_{r=1}^{R}\|\mathbf{G}_r\|_*+\left \langle \Lambda _4,\mathcal{X}-\mathcal{G}\times_3\mathbf{C} \right \rangle+\frac{\beta_4}{2}\left \| \mathcal{X}-\mathcal{G}\times_3\mathbf{C}\right \|_\text{F}^2,
\end{equation}
where $\mathcal{G}(:,:,r)=\mathbf{G}_r,r=1,\cdots,R$.  
The optimization problem (\ref{equation:update G}) can be equivalent to
\begin{equation}
\min_{\mathbf{G}_{r,r=1,\cdots,R}}  \lambda_{\mathcal{G}}\sum_{r=1}^{R}\|\mathbf{G}_r\|_*+\frac{\beta_4}{2}\sum_{r=1}^{R}\|\mathbf{G}_r-\mathcal{M}(:,:,r)\|_\text{F}^2,
\end{equation}
where $\mathcal{M}=(\mathcal{X}+\frac{\Lambda_4}{\beta_4})\times_3\mathbf{C}^{\text{T}} $.
The detailed solutions of $\mathcal{G}$ are concluded in Algorithm 1, where $\text{SVD}$ means singular value decomposition and the soft thresholding operator of $x$ is defined as:
\begin{equation}\label{equation:thr}
\text{Thr}_{\tau}(x)=\text{sgn}(x)\max(|x|-\tau,0).
\end{equation}

\begin{algorithm}
    \SetAlgoNoLine  
     \caption{Updating $\mathcal{G}$}
      \KwIn{ $\mathcal{X}\in\mathbb{R}^{I_1\times I_2\times I_3}; \Lambda_4,\beta_4;\lambda_{\mathcal{G}}; R$}
      \KwOut{$\hat{\mathcal{G}}$}
		\quad 1. update  $\tau=\frac{\lambda_{\mathcal{G}}}{\beta_4}$\\
		\quad 2.  $\mathcal{M}=(\mathcal{X}+\frac{\Lambda_4}{\beta_4})\times_3\mathbf{C}^{\text{T}} $\\
		\quad 3. \textbf{for} r=1 to R:\\
		 \qquad [$\mathbf{U},\bm\Sigma,\mathbf{V}$]=$\operatorname{SVD}$($\mathcal{M}(:,:,r)$)\\
		 \qquad $\hat{\bm\Sigma}=\text{Thr}_{\tau}(\bm\Sigma)$\\
		 \qquad $\mathbf{G}_{r}$=$\mathbf{U}\hat{\bm\Sigma}\mathbf{V}^{\operatorname{T}}$\\
		 \qquad $\hat{\mathcal{G}}(:,:,r)=\mathbf{G}_{r}$\\
		 \qquad \textbf{end for}\\
		    \end{algorithm}	
\subsection{Updating $\mathbf{C}$}
Fixing $\mathbf{C}$ in equation (\ref{equation:mvtf3}), we can obtain the following equation:
\begin{eqnarray}\label{update:C}
&&\min_{\mathbf{C}}\left \langle \Lambda _4,\mathcal{X}-\mathcal{G}\times_3\mathbf{C} \right \rangle+\frac{\beta_4}{2}\left \| \mathcal{X}-\mathcal{G}\times_3\mathbf{C}\right \|_\text{F}^2, \nonumber\\
&&~\text{s.~t.~} \mathbf{C}^{T}\mathbf{C}=\mathbf{I}_R.
\end{eqnarray}
Solving problem (\ref{update:C}) is equal to :
\begin{equation}
\max_{\mathbf{C}}\operatorname{trace}(\mathbf{G}_{(3)}(\Lambda_{4(3)}^{\text{T}}+\beta_4\mathbf{X}_{(3)}^{\text{T}})\mathbf{C})~\text{s.~t.~} \mathbf{C}^{\text{T}}\mathbf{C}=\mathbf{I}_R.
\end{equation}
Letting $\mathbf{M}=\mathbf{G}_{(3)}(\Lambda_{4(3)}^{\text{T}}+\beta_4\mathbf{X}_{(3)}^{\text{T}})$ and performing SVD on $\mathbf{M}$ as $\mathbf{M}=\mathbf{U}\bm\Sigma\mathbf{V}^{\text{T}}$, the solution of $\mathbf{C}$ can be obtained by
\begin{equation}\label{solution:C}
\mathbf{C}=\mathbf{V}\mathbf{U}^{\text{T}}.
\end{equation}
\subsection{Updating $\mathcal{X}$}
Fixing other variables to update $\mathcal{X}$, the optimization model (\ref{equation:mvtf3}) can be rewritten  as:
\begin{eqnarray}\label{update:X}
&&\min_{\mathcal{X}}\left \langle \Lambda _1,\mathcal{Y}-\mathcal{X} -\mathcal{S}-\mathcal{N}\right \rangle+\left \langle \Lambda _2,\mathcal{Z}-\mathcal{X} \right \rangle\nonumber\\
&&+\left \langle \Lambda _4,\mathcal{X}-\mathcal{G}\times_3\mathbf{C} \right \rangle
+\frac{\beta_{1}}{2}\left \| \mathcal{Y}-\mathcal{X}-\mathcal{S}-\mathcal{N} \right \|_\text{F}^2\nonumber\\
&&+\frac{\beta_2}{2}\left \| \mathcal{Z}-\mathcal{X} \right \|_\text{F}^2
+\frac{\beta_4}{2}\left \| \mathcal{X}-\mathcal{G}\times_3\mathbf{C} \right \|_\text{F}^2.
\end{eqnarray}
 Using the objective function in (\ref{update:X}) to derive the $\mathcal{X}$, we can obtain the solution of $\mathcal{X}$ as follows:
\begin{equation}\label{solution:X}
\mathcal{X}=\frac{\beta_1(\mathcal{Y}-\mathcal{S}-\mathcal{N})+\Lambda_1+\beta_2\mathcal{Z}+\Lambda_2+\beta_4(\mathcal{G}\times_3\mathbf{C} )-\Lambda_4}{\beta_1+\beta_2+\beta_4}.
\end{equation} 

\subsection{Updating $\mathcal{Z}$}
Similarly, the optimization model (\ref{equation:mvtf3}) with respect to $\mathcal{Z}$ can be rewritten as:
\begin{eqnarray}\label{equation:update:Z}
&&\min_{\mathcal{Z}}\left \langle \Lambda _2,\mathcal{Z}-\mathcal{X} \right \rangle+\left \langle \Lambda _3,\mathbf{L}-\mathfrak{D}(\mathcal{Z}) \right \rangle\nonumber\\
&&+\frac{\beta_2}{2}\left \| \mathcal{Z}-\mathcal{X} \right \|_\text{F}^2+\frac{\beta_3}{2}\left \| \mathbf{L} -\mathfrak{D}(\mathcal{Z})\right \|_\text{F}^2.
\end{eqnarray}
The solution of this optimization problem can be transformed into the solution of the following linear system:
\begin{equation}
(\beta_2\mathbf{I}+\beta_3\mathfrak{D}^{*}\mathfrak{D})\mathcal{Z}=\beta_2\mathcal{X}-\Lambda_2+\mathfrak{D}^{*}(\beta_3\mathbf{L}+\Lambda_3).
\end{equation}
Considering the block circulant structure of the operator $\mathfrak{D}^{*}\mathfrak{D}$, it can be transformed into the Fourier domain and fast calculated.  The fast computation of $\mathcal{Z}$ can be written as:
\begin{equation}\label{solution:Z}
\mathcal{Z}=\operatorname{ifftn}(\frac{\operatorname{fftn}(\mathcal{M})}{\beta_{2}\mathbf{1}+\beta_{3}(\operatorname{fftn}(\mathfrak{D}^{*}\mathfrak{D}))}),
\end{equation}
where $\mathcal{M}=\beta_2\mathcal{X}-\Lambda_2+\mathfrak{D}^{*}(\beta_3\mathbf{L}+\Lambda_3)
$; $\operatorname{fftn}$ and $\operatorname{ifftn}$ are 3D fast  Fourier transform and 3D fast  inverse Fourier transform, respectively.

\subsection{Updating $\mathbf{L}$}
To update $\mathbf{L}$, we fix all the rest variables in equation (\ref{equation:mvtf3}), and the sub-problem can be formulated as follows:
\begin{equation}\label{equation:update V}
\min_{\mathbf{L}}\lambda_{\text{TV}}\left \| \mathbf{L}\right \|_{1}+\left \langle \Lambda _3,\mathbf{L} -\mathfrak{D}(\mathcal{Z})\right \rangle+\frac{\beta_3}{2}\left \| \mathbf{L}-\mathfrak{D}(\mathcal{Z}) \right \|_\text{F}^2.
\end{equation}
It can be transformed into:
\begin{equation}\label{equation:update V1}
\min_{\mathbf{L}}\lambda_{\text{TV}}\left \| \mathbf{L}\right \|_{1}+\frac{\beta_3}{2}\left \| \mathbf{L}-(\mathfrak{D}(\mathcal{Z})-\frac{\Lambda_3}{\beta_3}) \right \|_\text{F}^2.
\end{equation}
The solution of $\mathbf{L}$ can be obtained by equation (\ref{equation:thr}) as follows:
\begin{equation}\label{solution:L}
\mathbf{L}=\text{Thr}_{\frac{\lambda_{\text{TV}}}{\beta_3}}(\mathfrak{D}(\mathcal{Z})-\frac{\Lambda_3}{\beta_3}). 
\end{equation}

\subsection{Updating $\mathcal{S}$}
By fixing all the rest variables in equation (\ref{equation:mvtf3}), we can update $\mathcal{S}$ by the sub-problem:
\begin{equation}\label{update:S}
\min_{\mathcal{S}}\lambda_{\mathcal{S}}\left \| \mathcal{S} \right \|_{1}+\left \langle \Lambda _1,\mathcal{Y}-\mathcal{X} -\mathcal{S}-\mathcal{N}\right \rangle+\frac{\beta_{1}}{2}\left \| \mathcal{Y}-\mathcal{X}-\mathcal{S}-\mathcal{N} \right \|_\text{F}^2.
\end{equation}
Similarly to solving $\mathbf{L}$ above, this optimization model can be solved by:
\begin{equation}\label{solution:S}
\begin{split}
\mathcal{S}=\text{Thr}_{\frac{\lambda_{\mathcal{S}}}{\beta_1}}(\mathcal{Y}-\mathcal{X}-\mathcal{N}+\frac{\Lambda_1}{\beta_1}).
\end{split}
\end{equation}
\subsection{Updating $\mathcal{N}$}
Fixing the other variables in equation (\ref{equation:mvtf3}), the optimization model to update $\mathcal{N}$ can be formulated as follows:
\begin{equation}
\min_{\mathcal{N}}\lambda_{\mathcal{N}}\left \| \mathcal{N} \right \|_{\text{F}}^{2}+\left \langle \Lambda _1,\mathcal{Y}-\mathcal{X} -\mathcal{S}-\mathcal{N}\right \rangle+\frac{\beta_{1}}{2}\left \| \mathcal{Y}-\mathcal{X}-\mathcal{S}-\mathcal{N} \right \|_\text{F}^2.
\end{equation}
It can be easily solved by:
\begin{equation}\label{solution:N}
\begin{split}
\mathcal{N}=\frac{\beta_1(\mathcal{Y}-\mathcal{X}-\mathcal{S})+\Lambda_1}{\beta_1+2\lambda_{\mathcal{N}}}.
\end{split}
\end{equation}
\subsection{Updating $\Lambda_1,\Lambda_2,\Lambda_3,\Lambda_4$}
The Lagrangian multiplier updating scheme can be as follows:
\begin{eqnarray}\label{solution:lambda}
&&\Lambda _{1}=\Lambda _{1}+\beta_1(\mathcal{Y}-\mathcal{X}-\mathcal{S}-\mathcal{N}\nonumber\\
&&\Lambda_2=\Lambda_2+\beta_2(\mathcal{Z}-\mathcal{X})\nonumber\\
&&\Lambda_3=\Lambda_3+\beta_3(\mathbf{L}-\mathfrak{D}(\mathcal{Z}))\nonumber\\
&&\Lambda_4=\Lambda_4+\beta_4(\mathcal{X}-\mathcal{G}\times_3\mathbf{C}).
\end{eqnarray}

We summarize the algorithm for the SRLRTR with partially orthogonal MVTF for HSI denoising in Algorithm 2.
\begin{algorithm}
   \SetAlgoNoLine  
     \caption{SRLRTR with partially orthogonal MVTF}
      \KwIn{Observed hyperspectral image $\mathcal{Y}\in\mathbb{R}^{I\times J\times K}$; the threshold of stopping iteration $\varepsilon$; $\beta_1,\beta_2,\beta_3,\beta_4$;  $\lambda_\mathcal{S}, \lambda_\mathcal{G}, \lambda_\mathcal{N},\lambda_{\text{TV}}$}
      \KwOut{clean image $\mathcal{X}$}
	    \textbf{Initialiazation}:\quad$\mathcal{X}=\mathcal{Y},\mathcal{Z}=\mathcal{N}=\mathcal{S}=0$;\quad$\Lambda_1=\Lambda_2=\Lambda_4=0$;\quad$\Lambda_3=0$;\\
        \textbf{While} not converged \textbf {do}\\
		\quad 1. update  $\mathcal {G}$  by Algorithm 1\\
		\quad 2. update  $\mathbf {C}$  via (\ref{solution:C}) \\
		\quad 3. update  $\mathcal {X}$  via (\ref{solution:X}) \\
		\quad 4. update $\mathcal{Z}$ via (\ref{solution:Z}) \\
		\quad 5. update $\mathbf{L}$ via (\ref{solution:L})\\
		\quad 6. update $\mathbf{S}$ via (\ref{solution:S})\\
		\quad 7. update $ \mathcal N $ via (\ref{solution:N})\\
		\quad 8. update dual variables by (\ref{solution:lambda})\\	
		\textbf{ End while} \\
		    \end{algorithm}
		    
 The convergence condition is reached when the relative error between two successive tensors $\mathcal{X}$  is smaller than a threshold, i.e. ${\left \| \mathcal{X}^{(k)}-\mathcal{X}^{(k+1)} \right \|_\text{F}^2}/{\left \| \mathcal{X}^{(k+1)} \right \|_\text{F}^2}\leq \varepsilon $  where $\mathcal{X}^{(k)}$ is the recovered image in  $k$-th iteration and $\varepsilon$ is the threshold. 		   	    

\subsection{Computational complexity analysis}

 The main complexity of the proposed algorithm comes from the updating of $\mathcal{G}$. For a hyperspectral image $\mathcal{X}\in\mathbb{R}^{I\times J\times K}$,  the most time-consuming part is from calculating $\mathcal{G}$, which needs to compute the SVD of $\mathcal{M}_{(:,:,r)}\in \mathbb{R}^{I\times J}, r=1,\cdots,R$. Assuming $I=J$, the computational complexity is $O(I^3)$.
 Therefore, the total complexity is $O(PRI^{3})$, where $P$ is the number of iterations.

\section{Experiments}
In this section,  two simulated datasets and two real datasets are used in the experiments to testify the effectiveness of the proposed algorithm. The reflectivity values of all pixels in the hyperspectral image are normalized between 0 and 1 before the experiment.
\subsection{Datasets Descriptions}
\begin{figure*}[h!]
    \centering
        \subfigure[WDC]{\label{fig:WDC}\includegraphics[width=0.24\hsize]{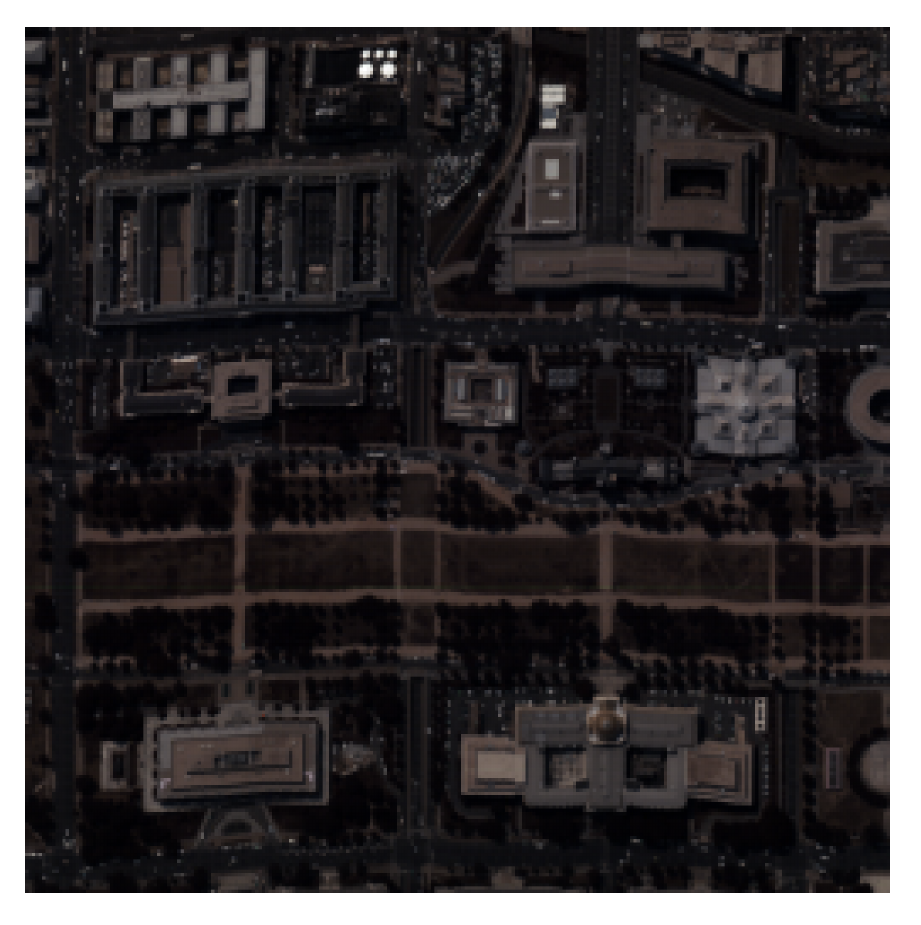}}  
        \subfigure[PaviaU]{\label{fig:PaviaU}\includegraphics[width=0.24\hsize]{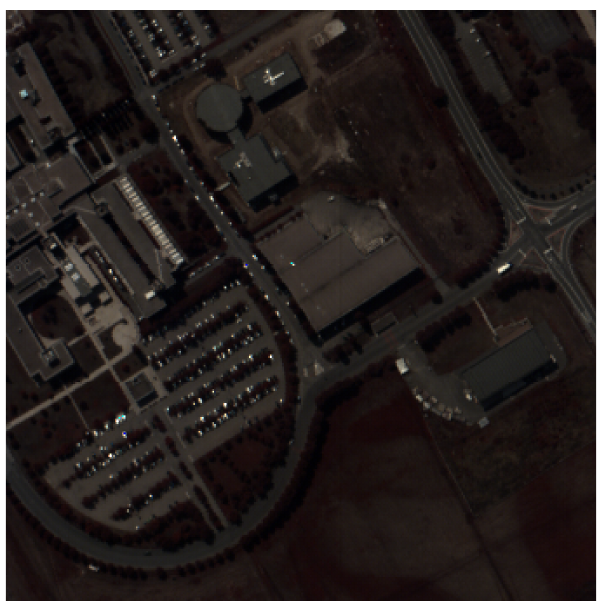}} 
         \subfigure[EO-1]{\label{fig:EO01}\includegraphics[width=0.24\hsize]{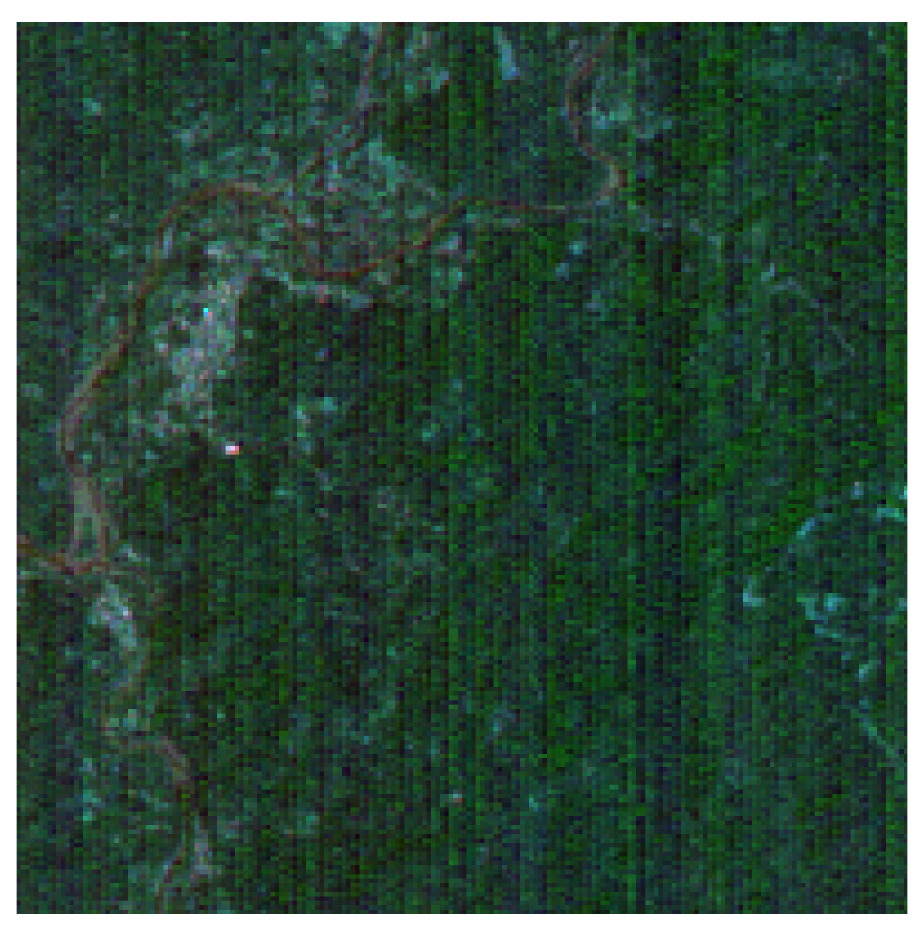}}      
         \subfigure[Urban]{\label{fig:Urban}\includegraphics[width=0.24\hsize]{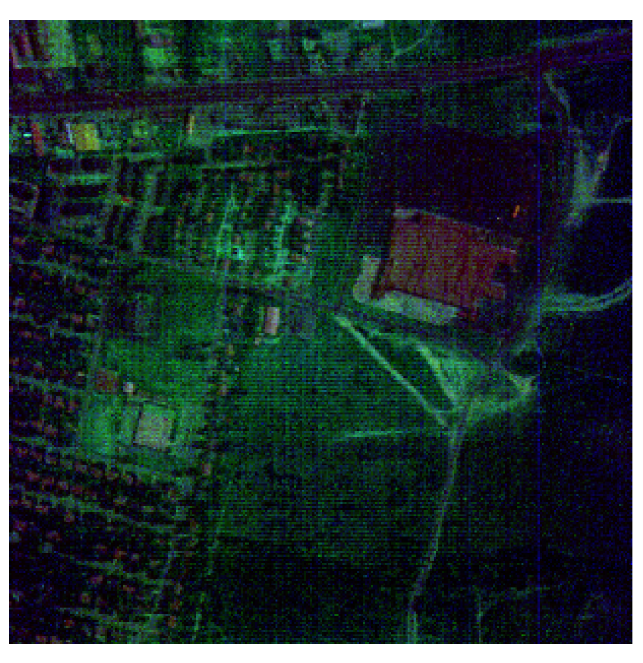}}      
           \caption{Testing HSI datasets}
    \label{fig:1}
\end{figure*}

\textbf{1.  Washington DC Mall (WDC)}:  The Washington DC Mall dataset\footnote[1]{https://engineering.purdue.edu/~biehl/MultiSpec/hyperspectral.html}  is provided with the permission of Spectral Information Technology Application Center of Virginia. The size of each scene is $1208\times307$. The dataset includes 210 bands in the spectral range from 400 nm to 2400 nm.  We use the 191 bands by removing bands where the atmosphere is opaque. And $256 \times 256 $ pixels are selected from each image, forming the HSI with the size of $256 \times 256 \times 191$. The false color image (R: 29, G: 19, B: 9) of WDC is shown in Fig. \ref{fig:WDC} .

\textbf{2. Pavia University (PaviaU)}: This scene\footnote[2]{https://rslab.ut.ac.ir/data}  is acquired by ROSIS sensor that covers the Pavia, northern Italy. The spatial size of the PaviaU is $610\times 340$ and there are 103 channels, ranging from 430 nm to 860 nm.  And $300\times 300$ pixels are considered, forming the HSI with the size of $300\times 300\times 103$. The false color image (R: 29, G: 19, B: 9) of PaviaU is shown in Fig. \ref{fig:PaviaU}.


\textbf{3. EO-1 Hyperion datasets (EO-1)}: The EO-1 Hyperion hyperspectral dataset\footnote[3]{https://www.usgs.gov/centers/eros/science/} covers an agricultural area of the state of Indiana, USA. It contains $1000\times 1000$ pixels and 242 bands spanning 350nm-2600nm. After removing water absorption bands, 166 bands are retained, and  $200\times 200$ pixels are selected from original scene, forming the HSI with size of $200\times 200\times 166$. It is mainly contaminated by Gaussian noise, stripes and dead lines.The false color image (R: 2, G: 132, B: 136) of EO-1 is shown in Fig. \ref{fig:EO01}. 

\textbf{4. HYDICE Urban Dataset (Urban)}: This scene is acquired from the real Hyperspectral Digital Imagery Collection Experiment (HYDICE)~\cite{mitchell1995hyperspectral}.  It contains $307\times 307$ pixels and there are 210 wavelengths ranging from 400 nm to 2500 nm. It is mainly contaminated by stripes and dead lines. The false color image (R: 2, G: 109, B: 207) of Urban is shown in Fig. \ref{fig:Urban}.

\begin{table*}[t]
\caption{Quantitative comparison of different algorithms for PaviaU and WDC denoising . }
\begin{center}
\resizebox{\textwidth}{!}{
\begin{tabular}{|c|c|c|c|c|c|c|c|c|c|c|c|c|c|}
\hline
 Cases                 &    Datasets  & measure indexes    & 3DATVLR & RSLRNTF & LRTDTV & SSTVLRTF & NLRCPTD&SNLRSF&TVLASDA&HyRes&BM4D& SRLRTR  \\ \hline
\multirow{8}{*}{Case 1}  &  & MPSNR (dB)&27.12& 19.26& 29.01& 20.93& 22.13&22.05 &26.65 &20.21 & 21.15&\textbf{29.17}\\ \cline{3-13}
 & & MSSIM& 0.73 & 0.44&\textbf{0.79} &0.51 & 0.57&0.69 &0.64 &0.62 &0.56&0.75 \\ \cline{3-13}
 & PaviaU& ERGAS& 172.50&424.75 &142.19 &325.18 &387.03 &5.1E+3 &197.91 &384.57 &344.21&\textbf{141.38} \\ \cline{3-13}
 & & CPU time (sec)& 217.04&723.36 &179.96 &729.27 &1.8E+4 &794.33 &391.09 &\textbf{3.42 }&549.27&245.45\\ \cline{2-13}
 &  & MPSNR (dB)& 26.07&19.20 &27.98 &21.44 &19.30 & 21.43&27.188&19.404&20.005&\textbf{29.642}\\ \cline{3-13}
 & & MSSIM&0.664 &0.59 &0.78 &0.61 &0.56&0.68 &0.79 &0.62&0.49& \textbf{0.85}\\ \cline{3-13}
 & WDC& ERGAS& 202.10& 490.13&162.25 &380.48 &281.5 &5.5E+3 & 181.64&480.19&432.21&\textbf{141.34} \\ \cline{3-13}
 & & CPU time (sec)& 212.72 &631.34 & 327.80&1.34E+3 & 3.98E+4&601.76&332.35&\textbf{1.98}&1.11E+3&271.30\\ \cline{1-13}
\multirow{8}{*}{Case 2}  &  & MPSNR (dB)& 29.12&31.29 &31.12 &30.70 &33.04 &\textbf{35.76} &28.87 &33.78 &32.05&32.63\\ \cline{3-13}
 & & MSSIM& 0.80& 0.819&0.851&0.746 &0.882 & \textbf{0.924}& 0.731&0.886&0.854&0.823\\ \cline{3-13}
 & PaviaU& ERGAS&139.72 &111.53 & 108.186& 121.21&90.21 &1.06E+3 &148.75 &\textbf{85.64} &101.95&98.45 \\ \cline{3-13}
 & & CPU time (sec)& 180.88& 446.2&132.57 &657.66  &2.7E+4 &2.1E+3 &321.665 &\textbf{1.649}&361.01 &98.08\\ \cline{2-13}
  &  & MPSNR (dB)&28.04 &31.89 &30.27 &30.71  &31.94&\textbf{36.25} &29.21&34.47&29.72&33.56 \\ \cline{3-13}
 & & MSSIM&  0.76 &0.91 & 0.85&0.89  &0.80&\textbf{0.966}&0.857&0.947&0.853&0.929 \\ \cline{3-13}
 & WDC& ERGAS& 160.74& 104.13&124.26 &107.48 &126.2 &1.05E+3 &142.42 &\textbf{78.10}&132.68&87.58\\ \cline{3-13}
 & & CPU time (sec)&536.77 &888.73 &272.92 &1.05E+3 &4.24E+4&590.02&431.91&\textbf{5.541}&683.03 &283.91 \\ \cline{1-13}
 \multirow{8}{*}{Case 3}  &  & MPSNR (dB)& 32.30&26.06 &33.23 &32.733 &26.42 &30.35 &32.35 & 27.44&24.92 &\textbf{36.75}\\ \cline{3-13}
 & & MSSIM& 0.89&0.709 &0.901 &0.825 &0.77 &0.875 &0.894 &0.80 &0.537 &\textbf{0.936} \\ \cline{3-13}
 & PaviaU& ERGAS&99.83 &207.99&87.02 &109.89  &181.37 &1.9E+3 &103.47 &185.12 &253.36&\textbf{63.12}\\ \cline{3-13}
 & & CPU time (sec)&167.85&811.507 &129.46&572.934 &2.3E+4 &1.4E+3 &537.47 &\textbf{3.626} &747.35.92 &182.65 \\ \cline{2-13}
 &  & MPSNR (dB)& 32.23&26.05 &32.96 &33.96  &26.04&29.72 &32.91&26.88&23.81&\textbf{37.85} \\ \cline{3-13}
 & & MSSIM& 0.90& 0.82&0.92 &0.93  &0.79&0.90&0.93&0.85&0.62&\textbf{0.97}\\ \cline{3-13}
 & WDC& ERGAS&97.85 &208.55 &91.81 &85.78 & 185.3&2.11E+3 &99.57&196.58&279.44&\textbf{58.97} \\ \cline{3-13}
 & & CPU time (sec)&339.54 &894.08&260.91 &1.46E+3 & 3.63E+4&598.662&430.395&\textbf{4.94}&772.359&287.08\\ \cline{1-13}
 \multirow{8}{*}{Case 4}  &  & MPSNR (dB)& 32.27 &25.75 &33.45 &31.77 &26.07 &30.04 &32.28 &26.85 & 25.04& \textbf{36.56}\\ \cline{3-13}
 & & MSSIM&0.888 &0.693 &0.904 &0.819 &0.751 &0.879 &0.888 &0.849 &0.543 &\textbf{0.934}\\ \cline{3-13}
 & PaviaU&ERGAS&100/18 &213.79 &87.93 &112.32& 187.81&2.10E+3 &104.43 &196.75 &250.74 &\textbf{63.34} \\ \cline{3-13}
 & & CPU time (sec)& 209.25&717.91 &182.04 &762.24 &3.6E+4 &790.49 &386.89 &\textbf{4.110 }&526.23 &211.807\\ \cline{2-13}
 &  & MPSNR (dB)&32.089 &25.83 &32.76 &32.99 & 25.54&29.62 &32.76&26.85&23.75&\textbf{37.181} \\ \cline{3-13}
 & & MSSIM& 0.897&0.80 &0.91 &0.926 &0.76&0.902&0.936&0.849&0.617  &\textbf{ 0.971}\\ \cline{3-13}
 & WDC& ERGAS&101.104 &214.84 &93.68 &97.85 & 202.7&2.14E+3 &100.31&196.73&279.65&\textbf{52.55} \\ \cline{3-13}
 & & CPU time (sec)& 295.158& 979.73&263.30 &1.17E+3 &4.31E+4&556.23&427.82&\textbf{4.11}&774.98  &314.45 \\ \cline{1-13}
\end{tabular}}
\end{center}
\label{table:1}
\end{table*}%
\begin{figure*}[h!]
    \centering
        \subfigure[Clean Image]{\includegraphics[width=0.15\hsize]{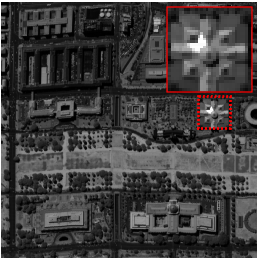}}  
        \subfigure[Noisy Image]{\includegraphics[width=0.15\hsize]{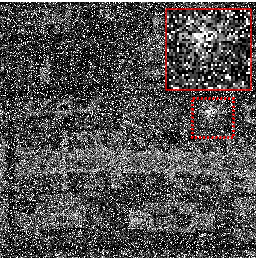}} 
        \subfigure[LRTDTV]{\includegraphics[width=0.15\hsize]{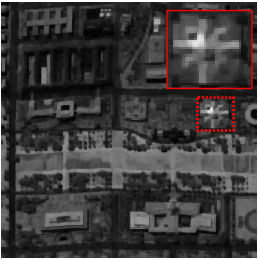}}           
        \subfigure[RSLRNTF]{\includegraphics[width=0.15\hsize]{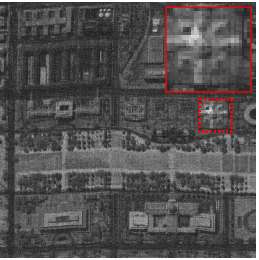}}      
         \subfigure[SNLRSF]{\includegraphics[width=0.15\hsize]{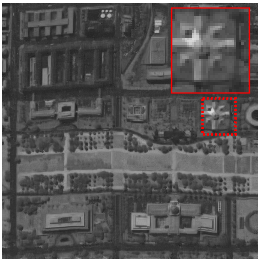}} 
                 \subfigure[3DATVLR]{\includegraphics[width=0.15\hsize]{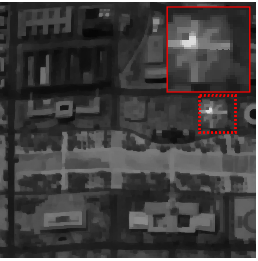}}\\
          \subfigure[NLRCPTD]{\includegraphics[width=0.15\hsize]{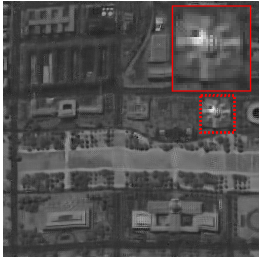}} 
        \subfigure[SSTVLRTF]{\includegraphics[width=0.15\hsize]{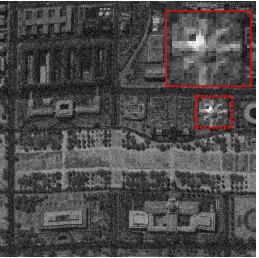}}
              \subfigure[TVLASDS]{\includegraphics[width=0.15\hsize]{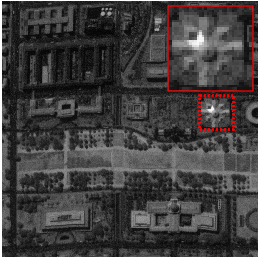}}
                \subfigure[HyRes]{\includegraphics[width=0.15\hsize]{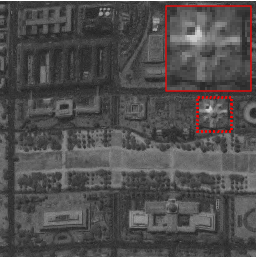}}
                  \subfigure[BM4D]{\includegraphics[width=0.15\hsize]{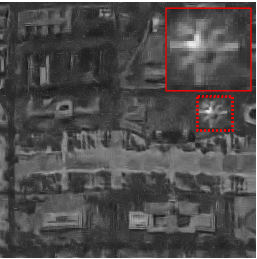}}
        \subfigure[SRLRTR]{\includegraphics[width=0.15\hsize]{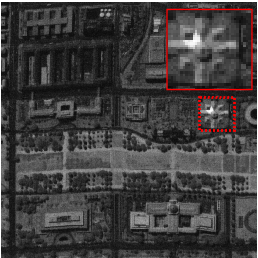}}     
           \caption{The recovered 121st spectral segment of the WDC data in case 1.}
    \label{fig:1}
\end{figure*}
 \begin{figure*}[h!]
    \centering
        \subfigure[Clean Image]{\includegraphics[width=0.15\hsize]{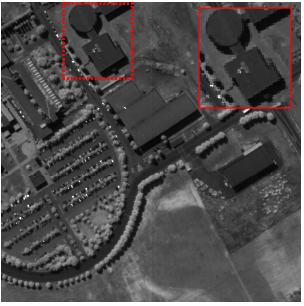}}  
        \subfigure[Noisy Image]{\includegraphics[width=0.15\hsize]{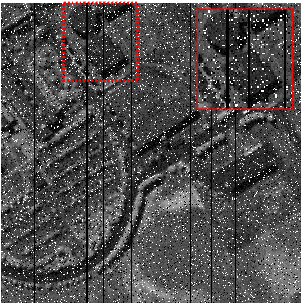}} 
        \subfigure[LRTDTV]{\includegraphics[width=0.15\hsize]{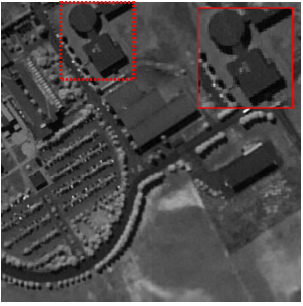}}            
        \subfigure[RSLRNTF]{\includegraphics[width=0.15\hsize]{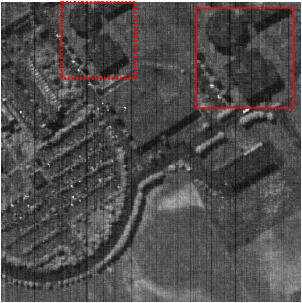}}      
         \subfigure[SNLRSF]{\includegraphics[width=0.15\hsize]{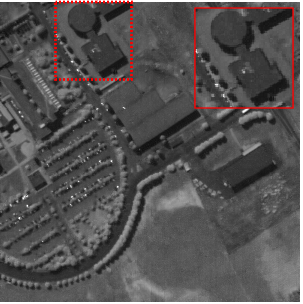}} 
                 \subfigure[3DATVLR]{\includegraphics[width=0.15\hsize]{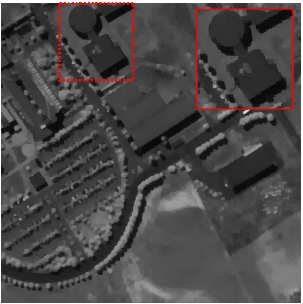}}\\
          \subfigure[NLRCPTD]{\includegraphics[width=0.15\hsize]{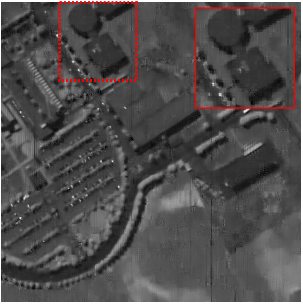}} 
        \subfigure[SSTVLRTF]{\includegraphics[width=0.15\hsize]{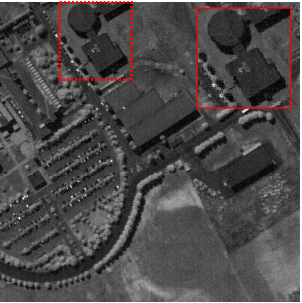}}
                      \subfigure[TVLASDS]{\includegraphics[width=0.15\hsize]{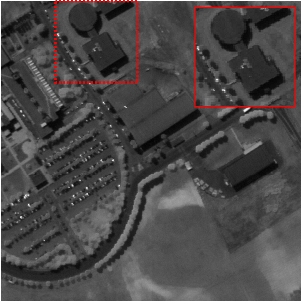}}
                \subfigure[HyRes]{\includegraphics[width=0.15\hsize]{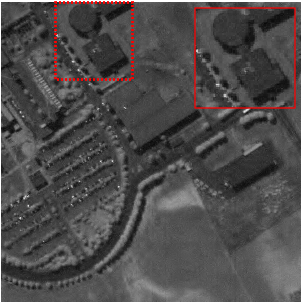}}
                  \subfigure[BM4D]{\includegraphics[width=0.15\hsize]{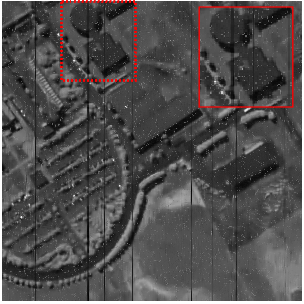}}
        \subfigure[SRLRTR]{\includegraphics[width=0.15\hsize]{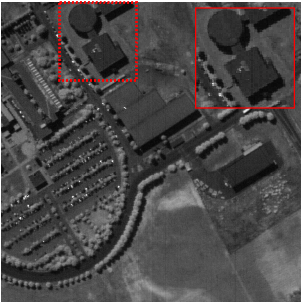}}      
           \caption{The recovered 85th spectral segment of the PaviaU data in case 3.}
    \label{fig:2}
\end{figure*}
\subsection{Competitive methods and measure metrics}
To compare with the proposed method, we select 9 state-of-the-art algorithms and one representative wavelet denoising algorithm. The detailed descriptions are as follows: 
\begin{itemize}
\item    SNLRSF~\cite{cao2019hyperspectral}: a novel subspace method based on nonlocal low-rank and sparse  constraints for mixed noise reduction of HSI.
\item  NLRCPTD~ \cite{xue2019nonlocal}: a method based on nonlocal low-rank regularized CP decomposition with rank determination for HSI denoising.
\item  3DATVLR ~\cite{ijgi7100412}: a  method based on 3D anisotropic total variation and low rank constraints  for mixed-noise removal of HSI.
\item  LRTDTV ~\cite{8233403}: a method based on total variation and low rank Tucker decomposition  for HSI mixed noise removal.
\item  RSLRNTF ~\cite{xiong2018hyperspectral}: a novel method based on  matrix-vector tensor factorization with low rank and nonnegative constrains for HSI denoising.
 \item SSTVLRTF~\cite{8367978}: a method based on spatial-spectral total variation and  low rank t-SVD factorization to remove mixed noise in HSI.
 \item TVLASDA~\cite{sun2018hyperspectral}: a method based on spectral difference-induced total variation and low rank approximation for HSI denoising.
  \item HyRes~\cite{rasti2017automatic}: a method based on sparse low-rank model, which automatically tunes parameter. 
 \item  BM4D~\cite{maggioni2012nonlocal}: a wavelet method based on nonlocal self-similarities, which achieves great success in nature image denoising.
\end{itemize}
The parameters of the comparison algorithms are set according to the values in its corresponding literatures. 

In addition,  to evaluate the performance quantitatively, four metrics are used to evaluate the image denoising quality, including MPSNR, MSSI,  ERGAS and CPU time. The details description are as follows: 
\begin{itemize}
\item $ \text{MPSNR}=\frac{1}{K}\sum_{k=1}^K \text{PSNR}_{k}$
\item $ \text{MSSIM}=\frac{1}{K}\sum_{k=1}^K \text{SSIM}_k$
\item $ \text{ERGAS}=\sqrt{\frac{1}{K}\sum_{k=1}^{K}{\left \| \hat{\mathcal{X}}(:,:,k)-\mathcal{X}(:,:,k) \right \|_{\text{F}}^2}/{\mu_k^2}}$
\end{itemize}
where $K$ is number of spectral bands. $\mu_k$ is the average value of $\mathcal{X}(:,:,k)$ and $\hat{\mathcal{X}}(:,:,k)$ is the recovered data.
 All the experiments are conducted using MatLab 2019b on a  computer with 2.4 GHz Quad-Core Intel Core i5 processor and 16 GB RAM.

\subsection{Experiments with simulated data}

 WDC and PaviaU hyperspectral images are used as simulated datasets in this experiment. To testify the proposed method, we add Gaussian noise, salt-and-pepper noise, dead line and strip noise to the simulated ground-truth datasets.   Four groups of experiments are performed, and the experimental settings are as follows:

	\textbf{Case 1:} Gaussian noise + impulse noise
	
	In this group of experiments, the noise intensities of different bands are the same. In each frequency band, the same zero mean Gaussian noise and the same percentage of impulse noise are added. The variance of Gaussian noise is 0.2, and the percentage of impulse noise is 0.2.
	
	\textbf{Case 2:} Gaussian noise +dead lines
	
	In this group of experiments, Gaussian noise is added to each band in the same way where the noise level is equal to 0.15. In addition, a dead line is added to the spectral segment from band 41 to band 100 with the number of stripes randomly chosen from 3 to 10.  The width of the stripes were randomly generated from one line to three lines.

	\textbf{Case 3:} Gaussian noise + impulse noise+dead lines
	
	In this group, the noise intensities of different bands are not the same. Gaussian noise with zero mean and  0.05 variance, is added to each  band. The percentage of impulse noise is 0.1. The dead lines are added randomly between 41 and 100 segments, the number of stripes is selected randomly from 3 to 10, and the width ranges randomly from one line to three lines.

	\textbf{Case 4:}  Gaussian noise + impulse noise+dead lines+stripe
	
	Based on the 3rd group of experiments, this group randomly adds stripe noise between band 101 and band 190 and the number of random stripe ranges from 20 to 40 for WDC data. For PaviaU, the dead lines are added randomly between 41 and 60 segments, the number of stripes is selected randomly from 3 to 10, and the width ranges randomly from one line to three lines. In addition, stripe noise is added between band 61 and band 100 when the number of random stripe ranges from 20 to 40.
\begin{figure}[h!]
\centering
\includegraphics[scale=0.35]{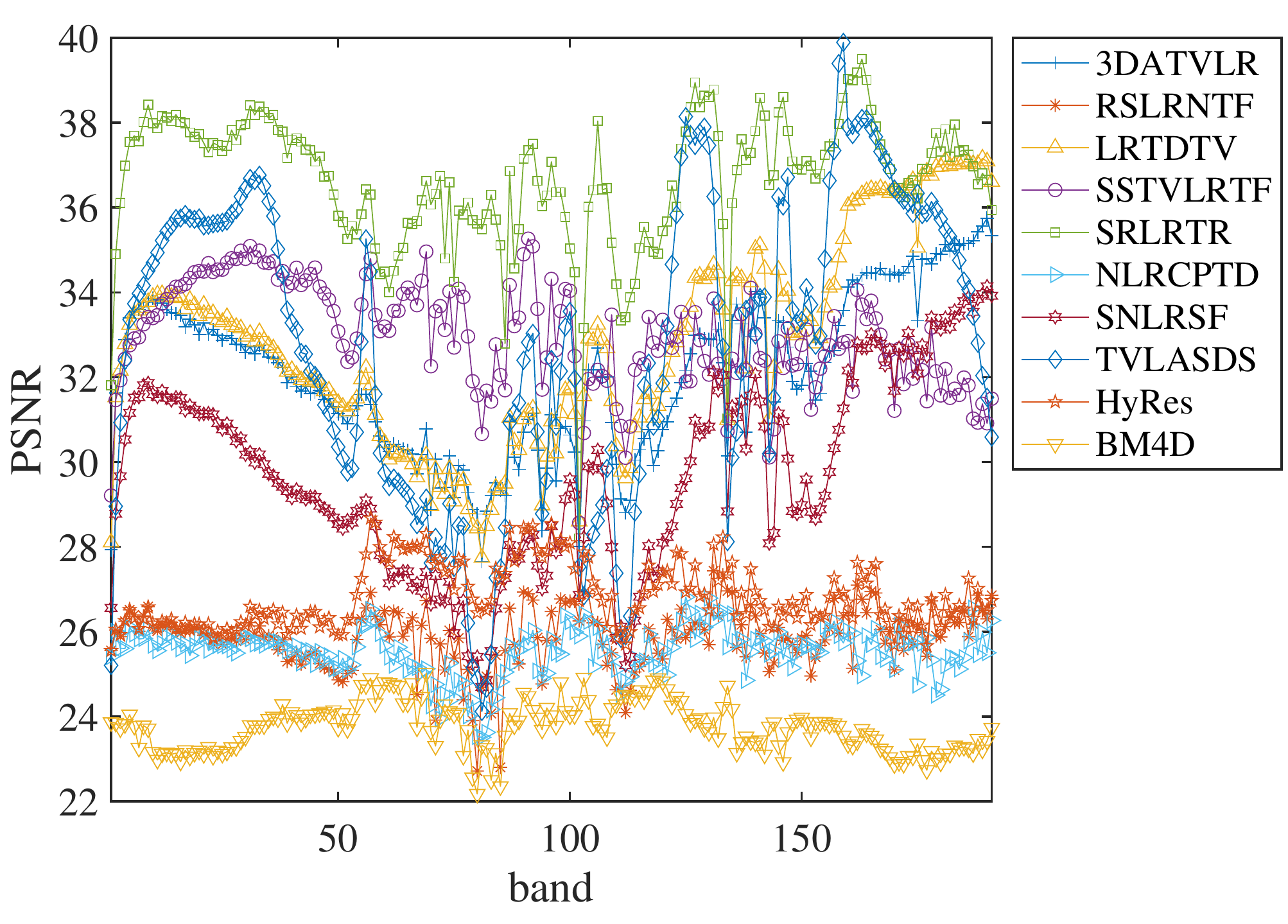}
\caption{The comparison performance of different methods in terms of PSNR in case 4 for WDC denoising.}
\label{fig:3}
\end{figure}

\begin{figure}[h!]
\centering
\includegraphics[scale=0.35]{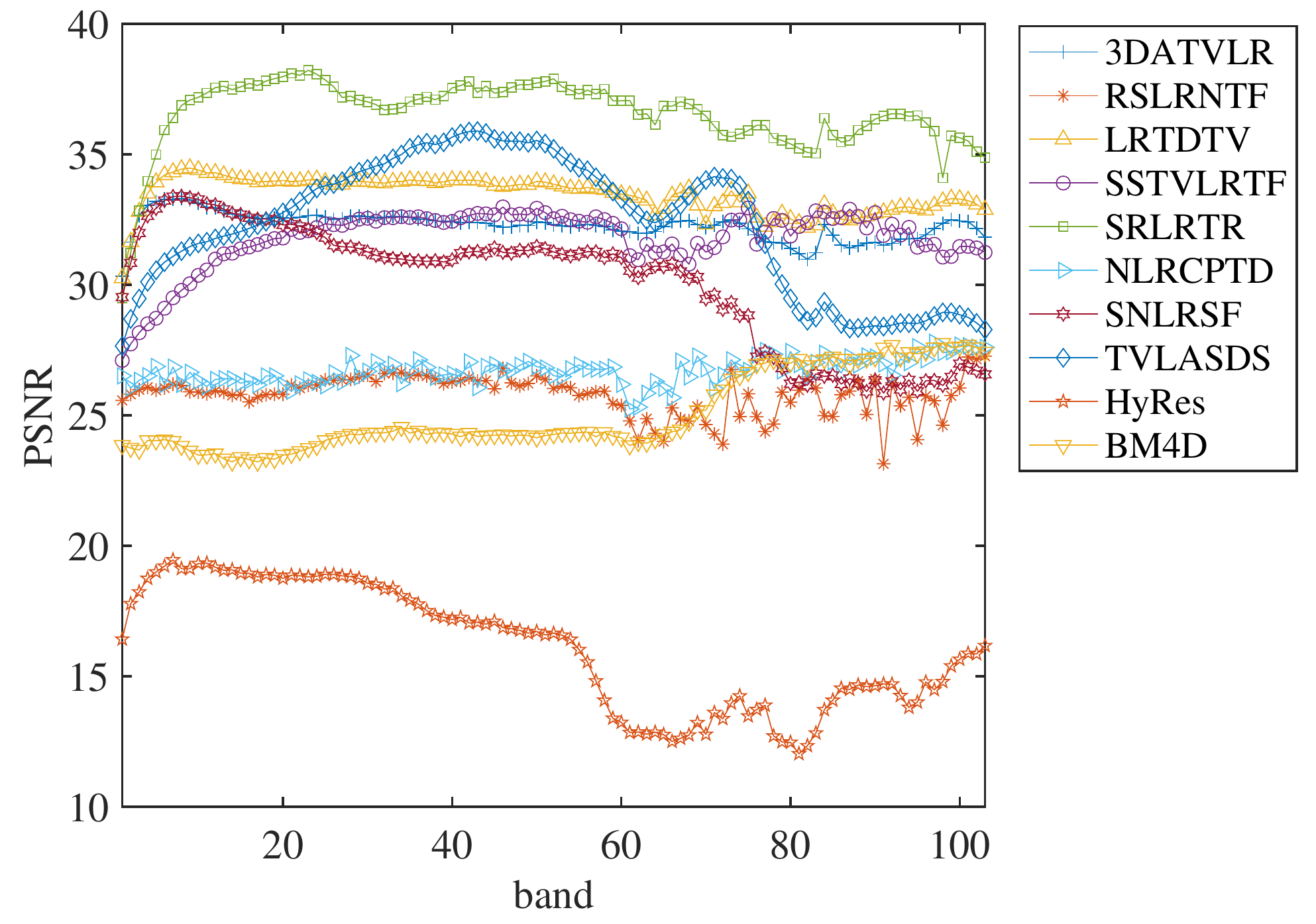}
\caption{The comparison performance of different methods in terms of PSNR in case 4 for PaviaU denoising.}
\label{fig:4}
\end{figure}
	
\begin{figure*}[h!]
    \centering  
                \subfigure[Noisy Image]{\includegraphics[width=0.15\hsize]{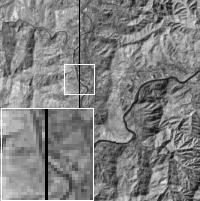}} 
        \subfigure[LRTDTV]{\includegraphics[width=0.15\hsize]{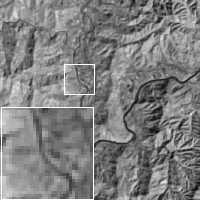}}          
        \subfigure[RSLRNTF]{\includegraphics[width=0.15\hsize]{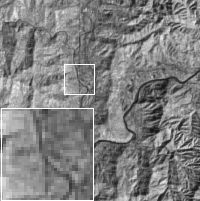}}      
         \subfigure[SNLRSF]{\includegraphics[width=0.15\hsize]{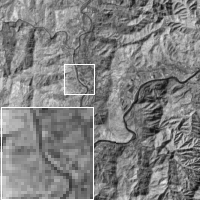}} 
                 \subfigure[3DATVLR]{\includegraphics[width=0.15\hsize]{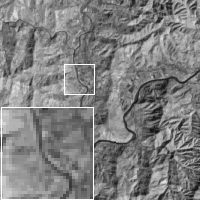}}
          \subfigure[NLRCPTD]{\includegraphics[width=0.15\hsize]{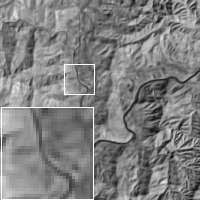}} 
        \subfigure[SSTVLRTF]{\includegraphics[width=0.15\hsize]{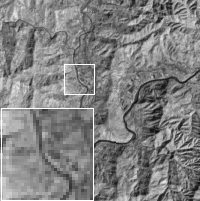}}
              \subfigure[TVLASDS]{\includegraphics[width=0.15\hsize]{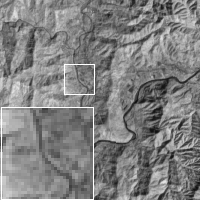}}
                \subfigure[HyRes]{\includegraphics[width=0.15\hsize]{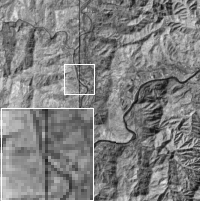}}
                  \subfigure[BM4D]{\includegraphics[width=0.15\hsize]{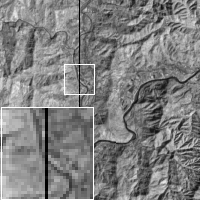}}
        \subfigure[SRLRTR]{\includegraphics[width=0.15\hsize]{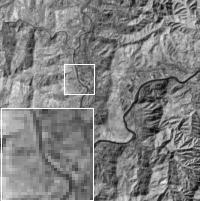}}      
           \caption{The recovered 68th spectral segment of the EO-01 data.}
               \label{fig:5}
\end{figure*}
\begin{figure*}[h!]
    \centering  
                \subfigure[Noisy Image]{\includegraphics[width=0.15\hsize]{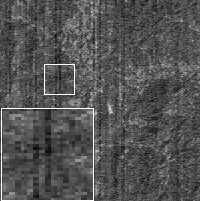}} 
        \subfigure[LRTDTV]{\includegraphics[width=0.15\hsize]{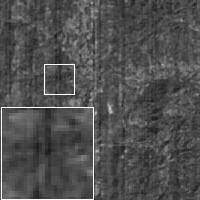}}          
        \subfigure[RSLRNTF]{\includegraphics[width=0.15\hsize]{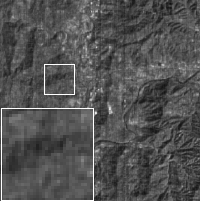}}      
         \subfigure[SNLRSF]{\includegraphics[width=0.15\hsize]{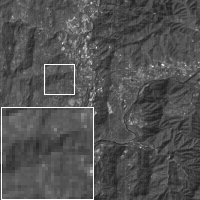}} 
                 \subfigure[3DATVLR]{\includegraphics[width=0.15\hsize]{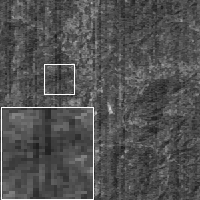}}
          \subfigure[NLRCPTD]{\includegraphics[width=0.15\hsize]{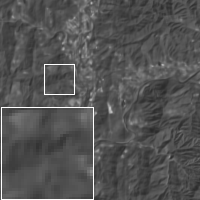}} 
        \subfigure[SSTVLRTF]{\includegraphics[width=0.15\hsize]{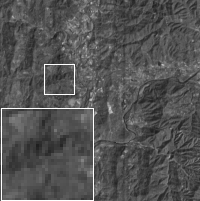}}
              \subfigure[TVLASDS]{\includegraphics[width=0.15\hsize]{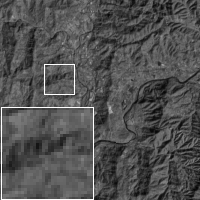}}
                \subfigure[HyRes]{\includegraphics[width=0.15\hsize]{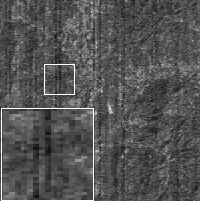}}
                  \subfigure[BM4D]{\includegraphics[width=0.15\hsize]{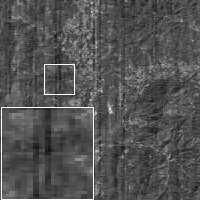}}
        \subfigure[SRLRTR]{\includegraphics[width=0.15\hsize]{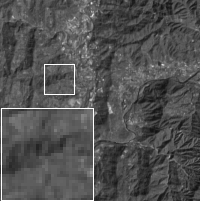}}      
           \caption{The recovered 96th spectral segment of the EO-01 data.}
               \label{fig:6}
\end{figure*}
\begin{figure*}[h!]
    \centering  
                \subfigure[Noisy Image]{\includegraphics[width=0.15\hsize]{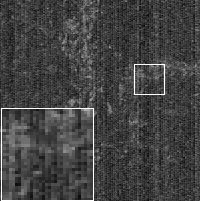}} 
        \subfigure[LRTDTV]{\includegraphics[width=0.15\hsize]{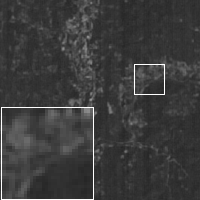}}          
        \subfigure[RSLRNTF]{\includegraphics[width=0.15\hsize]{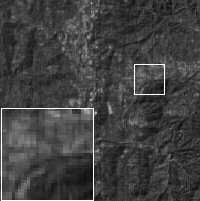}}      
         \subfigure[SNLRSF]{\includegraphics[width=0.15\hsize]{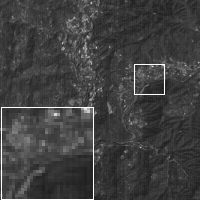}} 
                 \subfigure[3DATVLR]{\includegraphics[width=0.15\hsize]{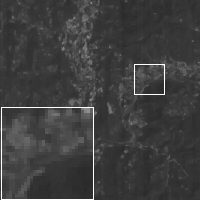}}
          \subfigure[NLRCPTD]{\includegraphics[width=0.15\hsize]{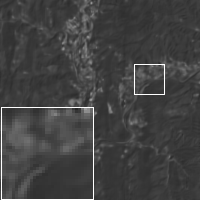}} 
        \subfigure[SSTVLRTF]{\includegraphics[width=0.15\hsize]{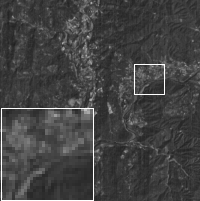}}
              \subfigure[TVLASDS]{\includegraphics[width=0.15\hsize]{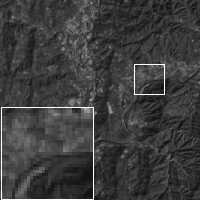}}
                \subfigure[HyRes]{\includegraphics[width=0.15\hsize]{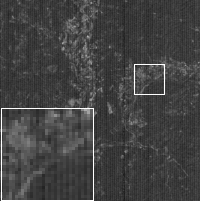}}
                  \subfigure[BM4D]{\includegraphics[width=0.15\hsize]{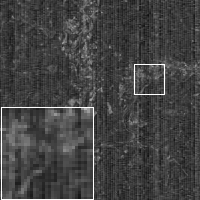}}
        \subfigure[SRLRTR]{\includegraphics[width=0.15\hsize]{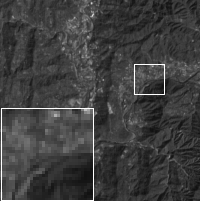}}      
           \caption{The recovered 132th spectral segment of the EO-01 data.}
               \label{fig:7}
\end{figure*}

 In the proposed method, we set the parameters $\lambda_\text{TV}=0.0002, \lambda_{\mathcal{S}}=0.02,\lambda_{\mathcal{N}}= 0.1,\lambda_{\mathcal{G}}=0.1$  and $R=5$ for WDC and PaviaU according to the highest value of MPSNR. The detailed information is put on the discussion part. 
 
Table \ref{table:1} shows the results of experiments based on simulated data in terms of MPSNR, MSSIM, ERGAS and CPU time. It can be observed, in most cases, the proposed algorithm is superior to all other methods in terms of the MPSNR, MSSIM and ERGA.  Specifically, our algorithm shows advantage on removing mixed Gaussian noise and impulse noise in case 1. It means that our algorithm can deal with Gaussian noise and impulse noise well. Similarly, in case 3 and case 4, when the dead line and stripe noise are added, our proposed method is also superior to existing state-of-art methods  in terms of recovery quality. This may imply the superiority of the proposed method in processing mixed noise in different conditions.  
However, in case 2 where it contains Gaussian noise and dead lines, SNLRSF, NLRCPTD and HyRes performs better. 
		
 To explain the recovery results more vividly, Fig. \ref{fig:1} and Fig. \ref{fig:2} present comparative results for the 121st band of WDC HSI in case 1 and the 85th band of PaviaU HSI in case 3, respectively.  Fig. \ref{fig:1} shows the recovered images by different methods on mixed Gaussian noise and impulse noise condition. We could observe that the resolution of recovered image in our proposed one is clearer than that of others. 
  In addition, Fig. \ref{fig:2} shows the recovered images on mixed Gaussian noise, impulse noise and dead lines condition. We could see that recovered images of  BM4D and RSLRTF still suffer from the dead lines noise. 

To give the details of results on each spectral segment, Fig. \ref{fig:3} and Fig. \ref{fig:4} provide the PSNR  of each spectral segment in case 4 for WDC and PaviaU data, respectively.  In case 4, we add different kinds of noises in different spectral segments. From Fig. \ref{fig:3}, we can see the PSNR values change smooth in the 1-41 spectral segments where Gaussian noise and impulse noise are added while the PSNR values fluctuate in 50-190 spectral segments where dead line and stripe noise are added.  In addition, for each scene, the recovery performance of the proposed one is better than that of others with respect to PSNR.
It indicates that the proposed method is superior in removing mixed noises. 
In Fig. \ref{fig:4}, the PSNR values of the proposed method is better than the others in almost every spectral segment.  It means that it can remove noise and retain the structural characteristics of HSI better. 

In conclusion, the proposed method performs better in the denoising of Gaussian noise and impulse noise. In the processing of stripe noise, it can also have a leading position among the compared methods.

\subsection{Experiments with real data}
We choose EO-01 and Urban datasets in this experiment. Both scenes are seriously contaminated by the Gaussian noise, stripes and dead lines. In this case, we set the parameters $\lambda_\text{TV}=0.00001, \lambda_{\mathcal{S}}=0.013,\lambda_{\mathcal{N}}= 0.1,\lambda_{\mathcal{G}}=0.1, R=2$ for Urban and EO-01data.

Fig. \ref{fig:5}, Fig. \ref{fig:6} and Fig \ref{fig:7} show the denoising resulting of EO-1 hyperspectral image in bands 68, 96 and 132.  It can be observed in Fig. \ref{fig:5},  the original data are heavily contaminated by stripes noise. Our proposed algorithm can recover the HSI while local details and structural information of the HSI are preserved. However, HyRes and BM4D fail to recover the HSI in stripes noise. This is mainly because they cannot model the sparse noise. Similarly, HyRes and BM4D performs badly in Fig. \ref{fig:6} and Fig. \ref{fig:7}. In addition, it can be observed that the proposed method obtains better performance in terms of image resolution.

\begin{figure*}[h!]
    \centering  
                \subfigure[Noisy Image]{\includegraphics[width=0.15\hsize]{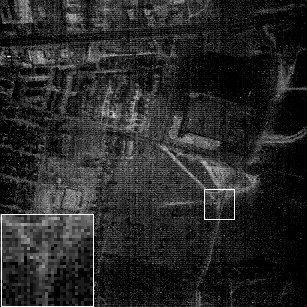}} 
        \subfigure[LRTDTV]{\includegraphics[width=0.15\hsize]{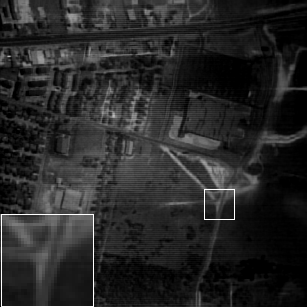}}          
        \subfigure[RSLRNTF]{\includegraphics[width=0.15\hsize]{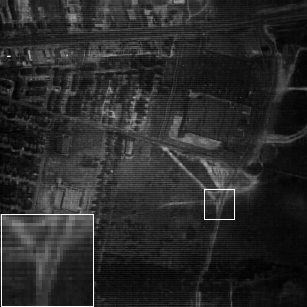}}      
         \subfigure[SNLRSF]{\includegraphics[width=0.15\hsize]{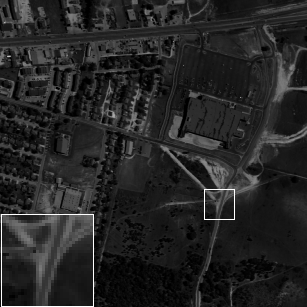}} 
                 \subfigure[3DATVLR]{\includegraphics[width=0.15\hsize]{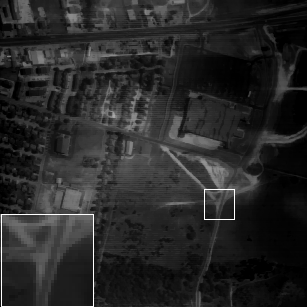}}
          \subfigure[NLRCPTD]{\includegraphics[width=0.15\hsize]{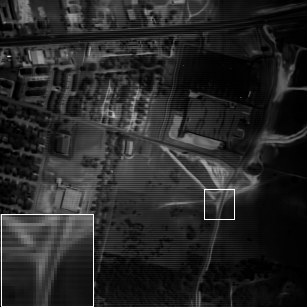}} 
        \subfigure[SSTVLRTF]{\includegraphics[width=0.15\hsize]{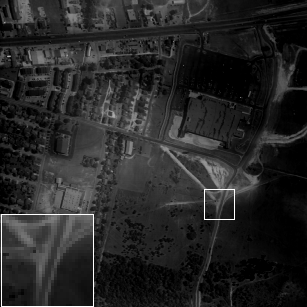}}
              \subfigure[TVLASDS]{\includegraphics[width=0.15\hsize]{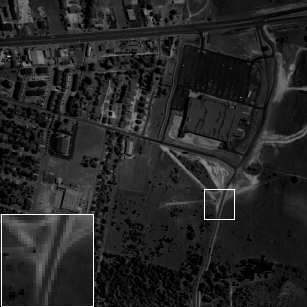}}
                \subfigure[HyRes]{\includegraphics[width=0.15\hsize]{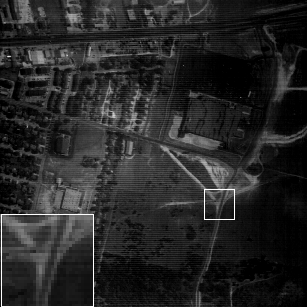}}
                  \subfigure[BM4D]{\includegraphics[width=0.15\hsize]{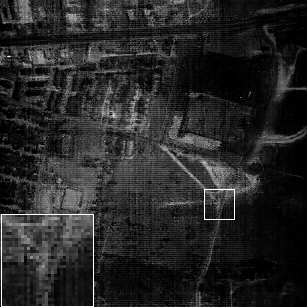}}
        \subfigure[SRLRTR]{\includegraphics[width=0.15\hsize]{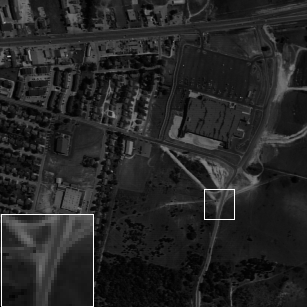}}      
           \caption{The recovered 109th spectral segment of the Urban data.}
               \label{fig:8}
\end{figure*}
\begin{figure*}[h!]
    \centering  
                \subfigure[Noisy Image]{\includegraphics[width=0.15\hsize]{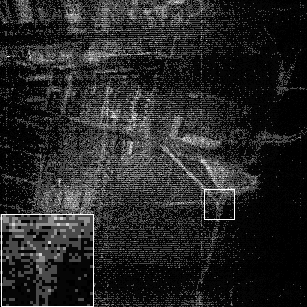}} 
        \subfigure[LRTDTV]{\includegraphics[width=0.15\hsize]{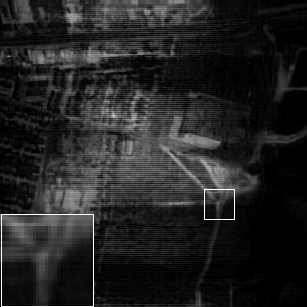}}          
        \subfigure[RSLRNTF]{\includegraphics[width=0.15\hsize]{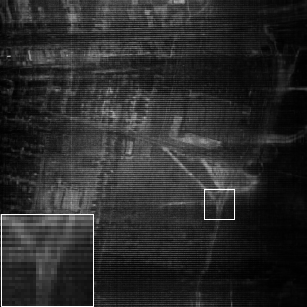}}      
         \subfigure[SNLRSF]{\includegraphics[width=0.15\hsize]{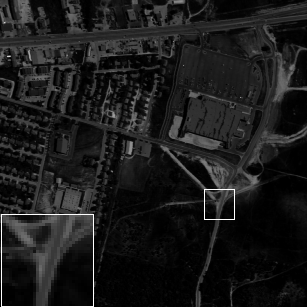}} 
                 \subfigure[3DATVLR]{\includegraphics[width=0.15\hsize]{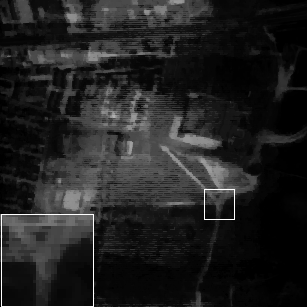}}
          \subfigure[NLRCPTD]{\includegraphics[width=0.15\hsize]{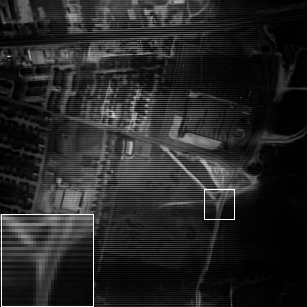}} 
        \subfigure[SSTVLRTF]{\includegraphics[width=0.15\hsize]{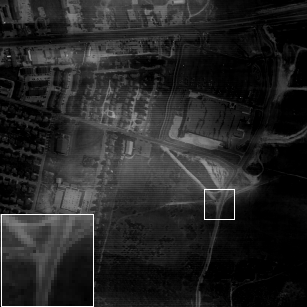}}
              \subfigure[TVLASDS]{\includegraphics[width=0.15\hsize]{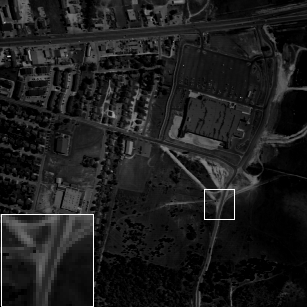}}
                \subfigure[HyRes]{\includegraphics[width=0.15\hsize]{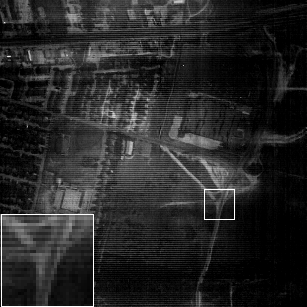}}
                  \subfigure[BM4D]{\includegraphics[width=0.15\hsize]{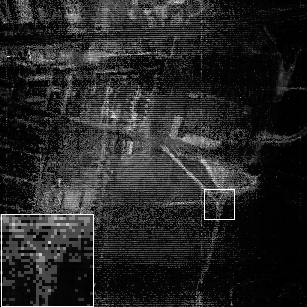}}
        \subfigure[SRLRTR]{\includegraphics[width=0.15\hsize]{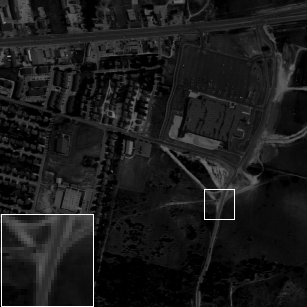}}      
           \caption{The recovered 151st spectral segment of the Urban data.}
               \label{fig:9}
\end{figure*}
\begin{figure*}[h!]
    \centering  
                \subfigure[Noisy Image]{\includegraphics[width=0.15\hsize]{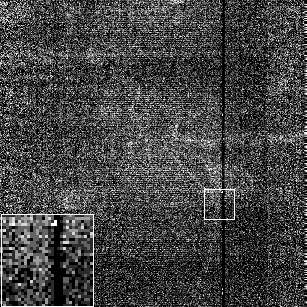}} 
        \subfigure[LRTDTV]{\includegraphics[width=0.15\hsize]{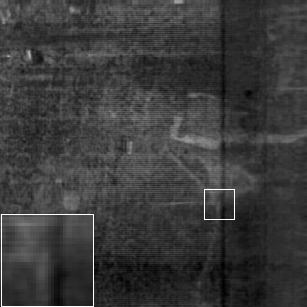}}          
        \subfigure[RSLRNTF]{\includegraphics[width=0.15\hsize]{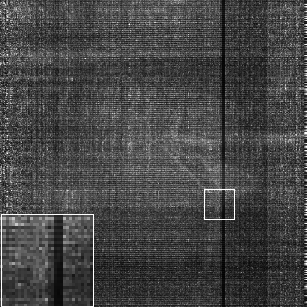}}      
         \subfigure[SNLRSF]{\includegraphics[width=0.15\hsize]{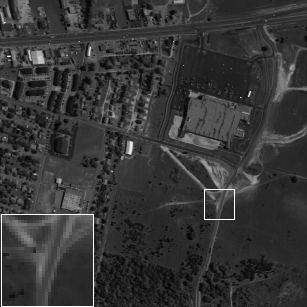}} 
                 \subfigure[3DATVLR]{\includegraphics[width=0.15\hsize]{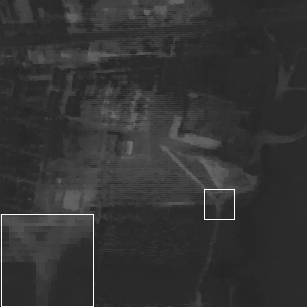}}
          \subfigure[NLRCPTD]{\includegraphics[width=0.15\hsize]{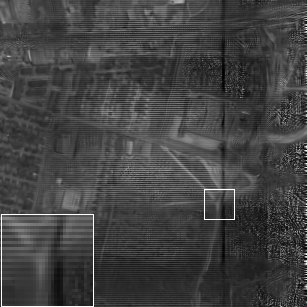}} 
        \subfigure[SSTVLRTF]{\includegraphics[width=0.15\hsize]{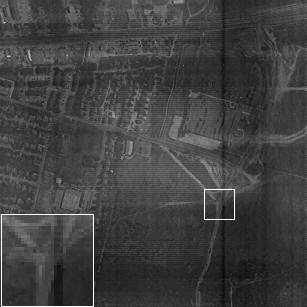}}
              \subfigure[TVLASDS]{\includegraphics[width=0.15\hsize]{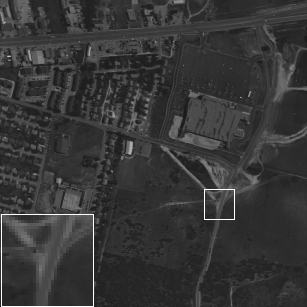}}
                \subfigure[HyRes]{\includegraphics[width=0.15\hsize]{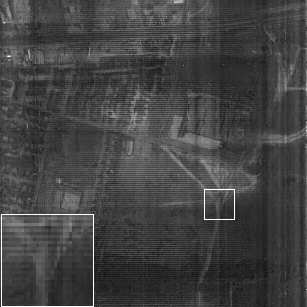}}
                  \subfigure[BM4D]{\includegraphics[width=0.15\hsize]{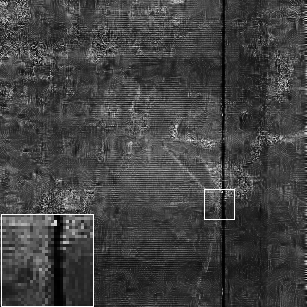}}
        \subfigure[SRLRTR]{\includegraphics[width=0.15\hsize]{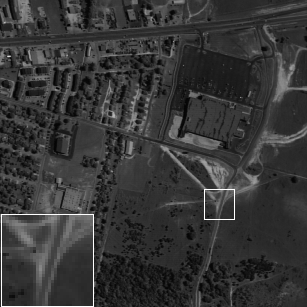}}      
           \caption{The recovered 208th spectral segment of the Urban data.}
               \label{fig:10}
\end{figure*}

 Fig. \ref{fig:8}, Fig. \ref{fig:9} and Fig. \ref{fig:10} show the 109th, 151st and 207th spectrum segments of Urban hyperspectral image after denoising.  Fig. \ref{fig:8} shows that the TV based methods including LRTDTV, 3DATVLR, SSTVLRTF, TVLASDS and our proposed one have a good recovery performance on stripe noise condition. Meanwhile, the nonlocal based methods such as SNLRSF and NLRCPTD can recover the image. However,  with the noise increase as shown in Fig. \ref{fig:9}, some TV based methods including LRTDTV, 3DATVLR and SSTVLRTF are over-smoothness to recover the HSI. Instead, SNLRSF, TVLASDS and our proposed can successfully recover the clean image from the severely noise condition. Specially, as shown in 
 Fig. \ref{fig:10}, the  scene is severely polluted by serious Gaussian noise, stripe noise and dead line, only SNLRSF, TVLASDS and our proposed method have a  good recovery performance.


\begin{table*}[h!]
\caption{ Comparison of different algorithms on Urban and EO-1 denoising in terms of CPU time . }
\begin{center}
\resizebox{\textwidth}{!}{
\begin{tabular}{|c|c|c|c|c|c|c|c|c|c|}
\hline
    Datasets  & measure indexes    & 3DATVLR & RSLRNTF & LRTDTV & SSTVLRTF   & SNLRSF& NLRCPTD&TVLASDS& SRLRTR  \\ \hline
  Urban data& CPU time (sec)& 730.26&1.68E+3 &558.40 &2.82E+3 & 1.91E+3&6.03E+4 &1.14E+3&620.03 \\ \cline{1-10}
 EO-1 data& CPU time (sec)& 128.38 &574.46&192.97 &1.38E+3 &805.01 & 5.78E+4&247&163.29 \\ \cline{1-10}
\end{tabular}}
\end{center}
\label{table:2}
\end{table*}%
\begin{figure*}[h!]
    \centering
        \subfigure[$\lambda_{\mathcal{N}}=0.01$]{\includegraphics[width=0.18\hsize]{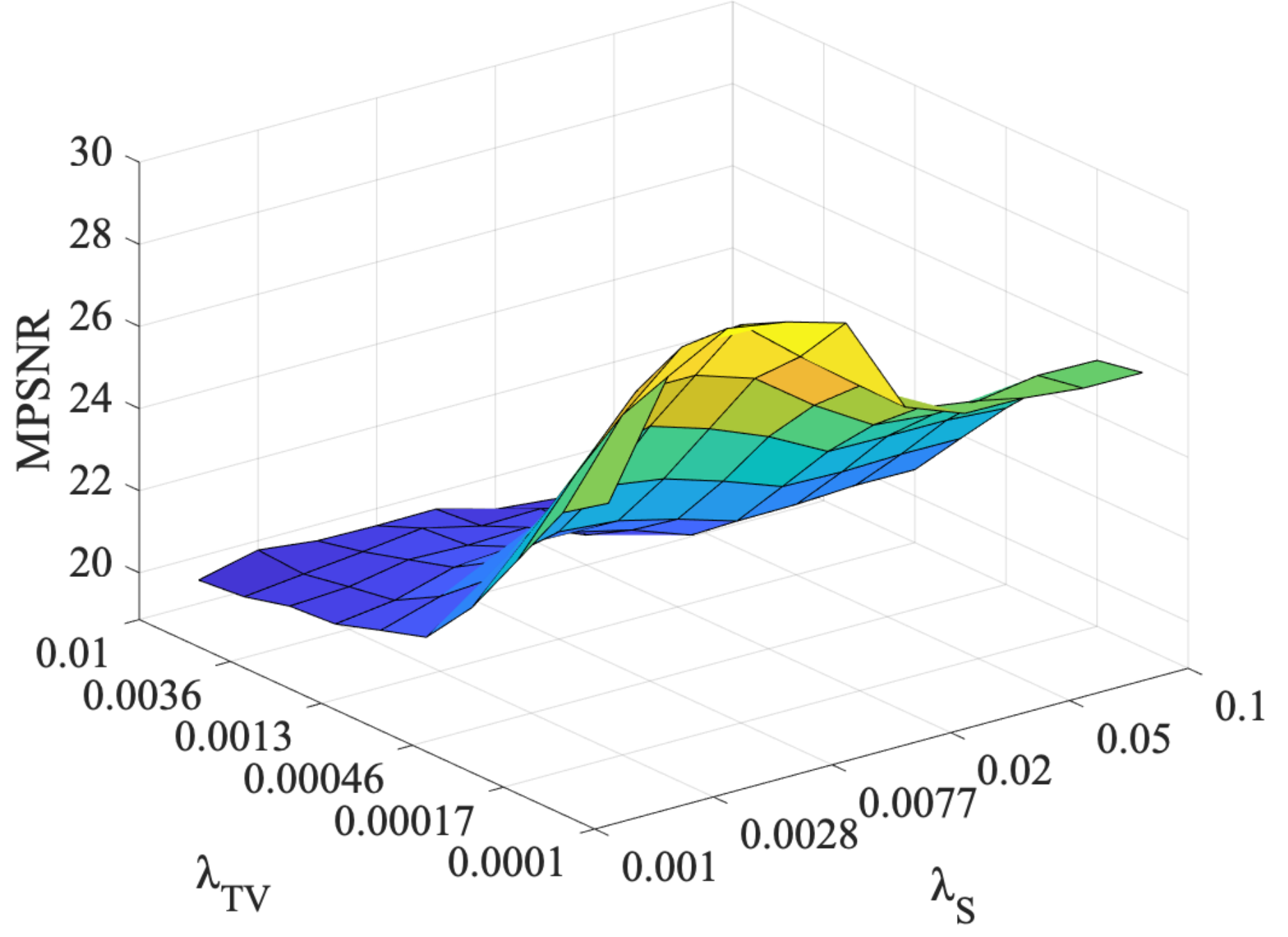}}  
        \subfigure[$\lambda_{\mathcal{N}}=0.018$]{\includegraphics[width=0.18\hsize]{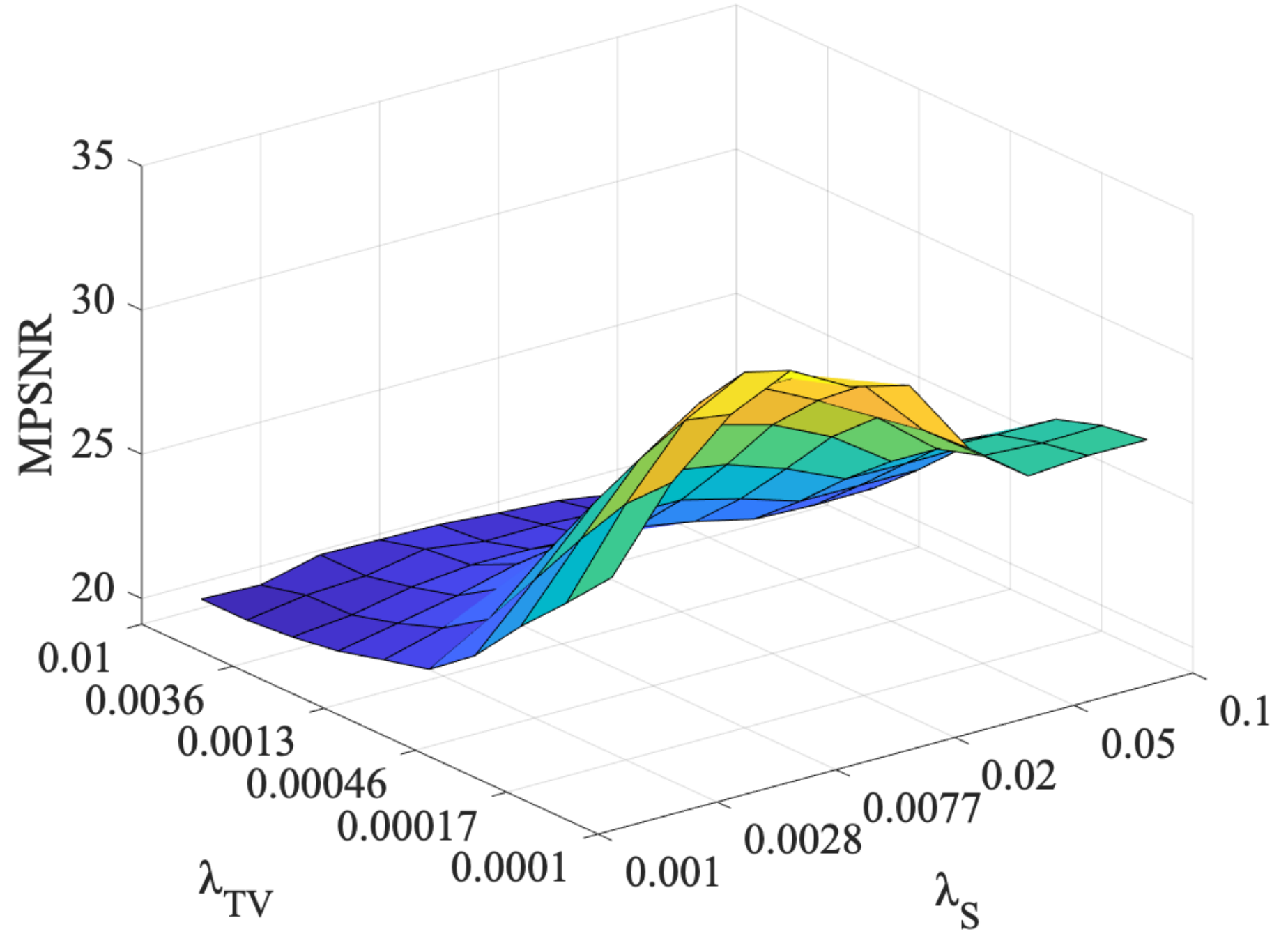}} 
        \subfigure[$\lambda_{\mathcal{N}}=0.032$]{\includegraphics[width=0.18\hsize]{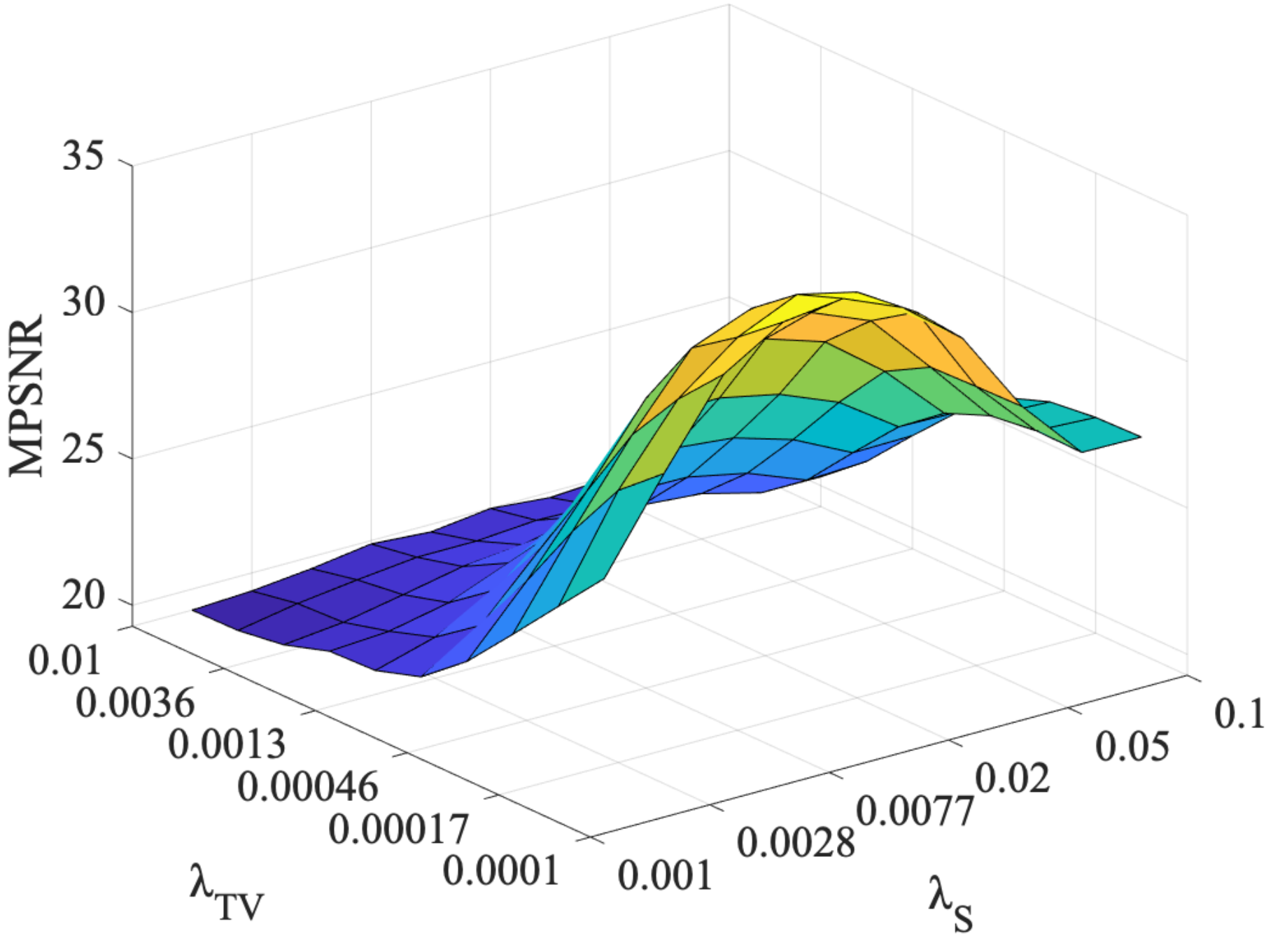}}           
        \subfigure[$\lambda_{\mathcal{N}}=0.056$]{\includegraphics[width=0.18\hsize]{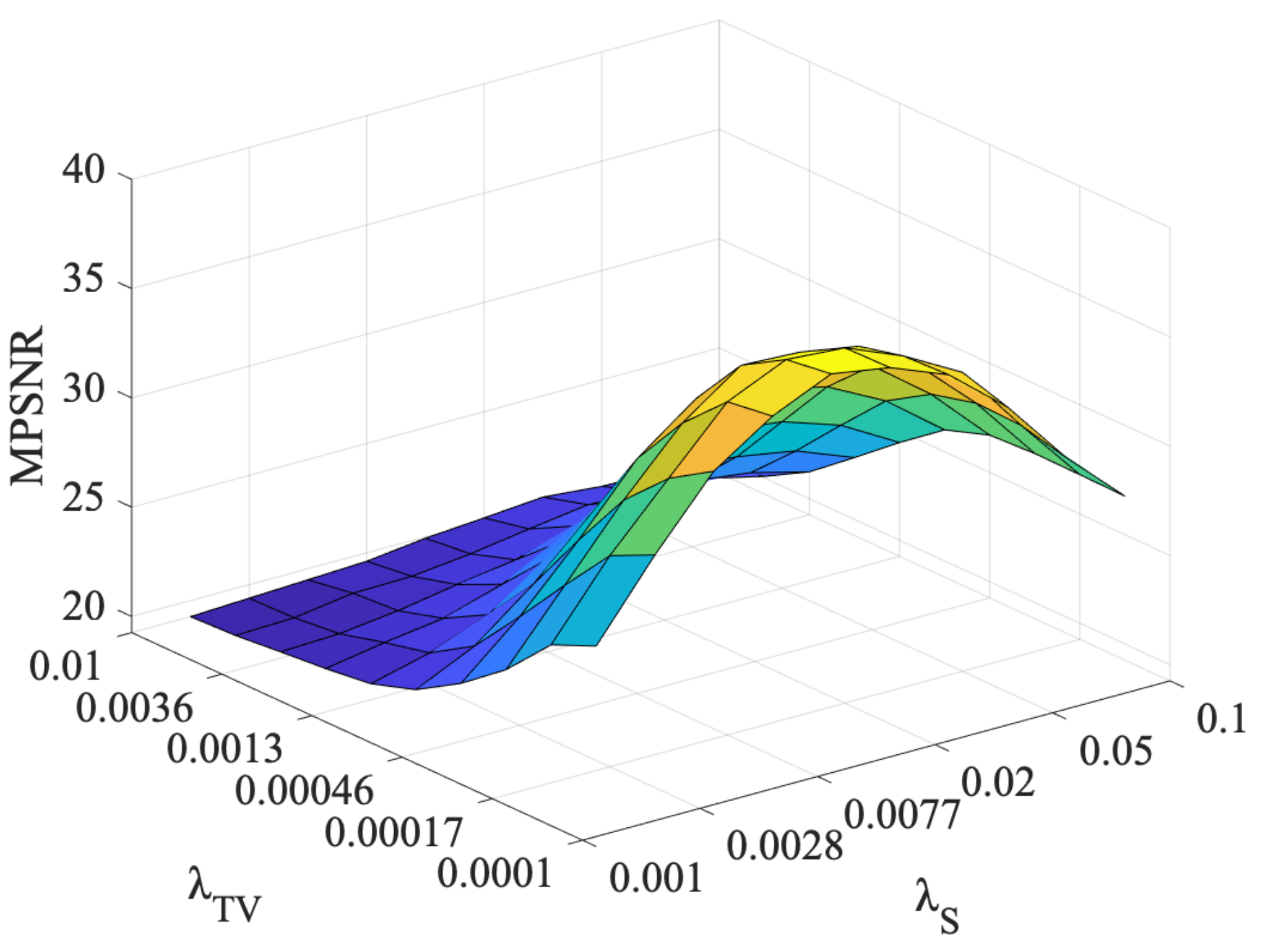}}      
         \subfigure[$\lambda_{\mathcal{N}}=0.1$]{\includegraphics[width=0.18\hsize]{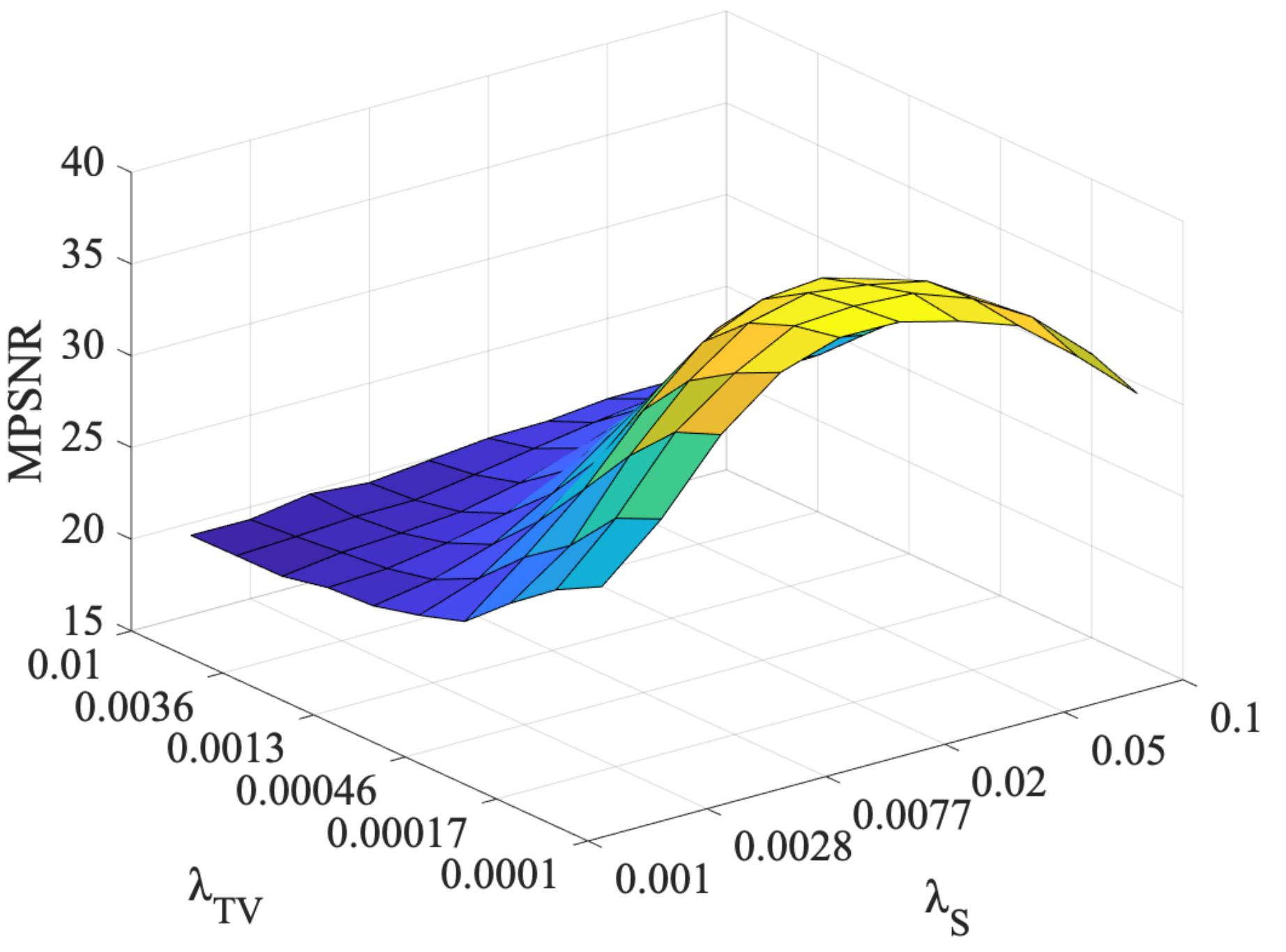}} 
           \caption{The impact of parameters $\lambda_{\text{TV}}$, $\lambda_{\mathcal{S}}$ and $\lambda_{\mathcal{N}}$.}
    \label{fig:parameters}
\end{figure*}
In addition, table \ref{table:2} shows the CPU time of  our algorithm and state-of-art algorithms on real data. It is noticed that due to the poor performance of the BM4D and HyRes methods in the above experiments, we do not consider the CPU time of them for comparison. 
As seen in table \ref{table:2}, 3DATVLR is the fastest one compared with others, followed by our proposed one. But the recovery performance of 3DATVLR is worse than that of SRLRTR. Meanwhile, compared  with TVLASDS and SNLRSF which have a good performance on severely noisy condition, the proposed one is faster than them. 
 It implies that the  partially orthogonal MVTF method can be more efficient and suitable to explore HSI information.

\section{Discussion}
\subsection{The difference between partially orthogonal MVTF and traditional MVTF }
We introduce a new smooth and robust low rank partially orthogonal MVTF method for  denoising of HSI.
The partially orthogonal MVTF is crucial to model the clean HSI for the proposed algorithm.  Compared with traditional MVTF including orthogonal MVTF~\cite{Lathauwer} and nonnegative MVTF~\cite{7784711}, the proposed one is different in two aspects. 

1. The traditional orthogonal MVTF in~\cite{Lathauwer} is defined as:
\begin{equation}
\mathcal{X}\approx \sum _{r=1}^{R}\mathbf{G}_r\circ \mathbf{c}_r= \sum _{r=1}^{R}\mathbf{D}_r\times_1\mathbf{A}_r\times_2\mathbf{B}_r\times_3\mathbf{c}_r,
\end{equation}
where $\mathbf{G}_r=\mathbf{A}_r\mathbf{D}_r\mathbf{B}_r^{\text{T}}$ is the SVD of $\mathbf{G}_r$ and $\mathbf{c}_r$ is under the unit-norm constraint. In this case, the matrices $\mathbf{A}_r$ and $\mathbf{B}_r$ are orthogonal. Compared with traditional orthogonal MVTF, the proposed one constrains $\mathbf{C}^{\text{T}}\mathbf{C}=\mathbf{I}_R$, $\mathbf{C}(:,r)=\mathbf{c}_r, r=1,\cdots,R$, resulting in partially orthogonal MVTF.  The partially orthogonal MVTF has a good  physical interpretation under the assumption of linear spectral mixture model for HSI, when $\mathbf{c}_r$ is considered as the $r$-th endmember spectrum, and endmember spectrums are uncorrelated. 


2. The nonnegative MVTF in~\cite{7784711} is denoted by:
\begin{eqnarray}
&&\mathcal{X}\approx \sum _{r=1}^{R}\mathbf{G}_r\circ \mathbf{c}_r= \sum _{r=1}^{R}\mathbf{A}_r\mathbf{B}_r^{\text{T}}\circ \mathbf{c}_r\nonumber\\
&& \text{s.~t.~} \mathbf{A}_r, \mathbf{B}_r,\mathbf{c}_r\geq 0,
\end{eqnarray}
where  $\mathbf{G}_r$ is of rank $L_r, r=1,\cdots,R$, and can be factorized as $\mathbf{A}_r\mathbf{B}_r^{\text{T}}$ with predefined rank $L_r$. In  HSI,  the abundance matrix is low rank.

This low rank model has shown its superiority on HSI unmixing~\cite{7784711}, where $\mathbf{c}_r$ is considered as an endmember, and  matrix $\mathbf{G}_r$ is the corresponding abundance map. In \cite{xiong2018hyperspectral}, RSLRNTF uses this model for HSI denoising. In these cases, the ranks $L_r$, $r=1, \cdots, R$ are required to be defined in advance. In practice, accurate determination of the rank of abundance matrix $\mathbf{G}_r$ of the $r$-th endmember is impossible. Therefore, we directly minimize the rank of each abundance matrix under the assumption that endmember spectrums are irrelevant, resulting in low rank partially-orthogonal MVTF. In addition, compared with RSLRNTF~\cite{xiong2018hyperspectral},  our proposed method shows a better recovery performance in experiments.

\subsection{The impact of parameters}
There are four parameters in equation (\ref{equation:mvtf}), but we only need to tune three parameters because the parameters represent the relative weights of different terms in objective function. In this case, we fix $\lambda_{\mathcal{G}}$ and analyze the impacts of parameters $\lambda_{\mathcal{S}},\lambda_{\mathcal{N}},\lambda_{\text{TV}}$.  In addition, the parameter $R$ in our algorithm has its physical meaning, which represents the number of endmembers (also called pure materials). In the following, we will take the experiments in case 4 on WDC datasets as an example and address how to choose these parameters.


1. The impact of parameters $\lambda_{\text{TV}}$, $\lambda_{\mathcal{S}}$ and $\lambda_{\mathcal{N}}$:
For tuning the parameters $\lambda_{\text{TV}}$, $\lambda_{\mathcal{S}}$ and $\lambda_{\mathcal{N}}$, we set $\lambda_{\mathcal{G}}=0.1$. The other three parameters are related to the weights of TV regularizer, sparse noise (i.e., impulse noise, stripes, and dead lines),  and Gaussian noise, respectively.  Fig. \ref{fig:parameters} shows the MPSNR values as a function with respect to $\lambda_{\text{TV}}$ and $\lambda_{\mathcal{S}}$ for the proposed algorithm with $\lambda_{\mathcal{N}}$ chosen from the set \{0.01, 0.018, 0.032, 0.056, 0.1\}.  $\lambda_{\text{TV}}$ is selected from the set \{0.0001, 0.00017, 0.00046, 0.0013, 0.0036, 0.01\} and $\lambda_{\mathcal{S}}$ is selected from the set \{0.001, 0.0028, 0.0077, 0.02, 0.05, 0.1\}. It is obvious that the smaller $\lambda_{\text{TV}}$ and the larger $\lambda_{\mathcal{N}}$, the value of MPSNR  becomes larger. When $\lambda_{\mathcal{S}}=0.02$, the proposed method can reach the peak of MPSNR.
\begin{figure*}[h!]
    \centering
        \subfigure[$\lambda_{\mathcal{N}}=0.01$]{\includegraphics[width=0.18\hsize]{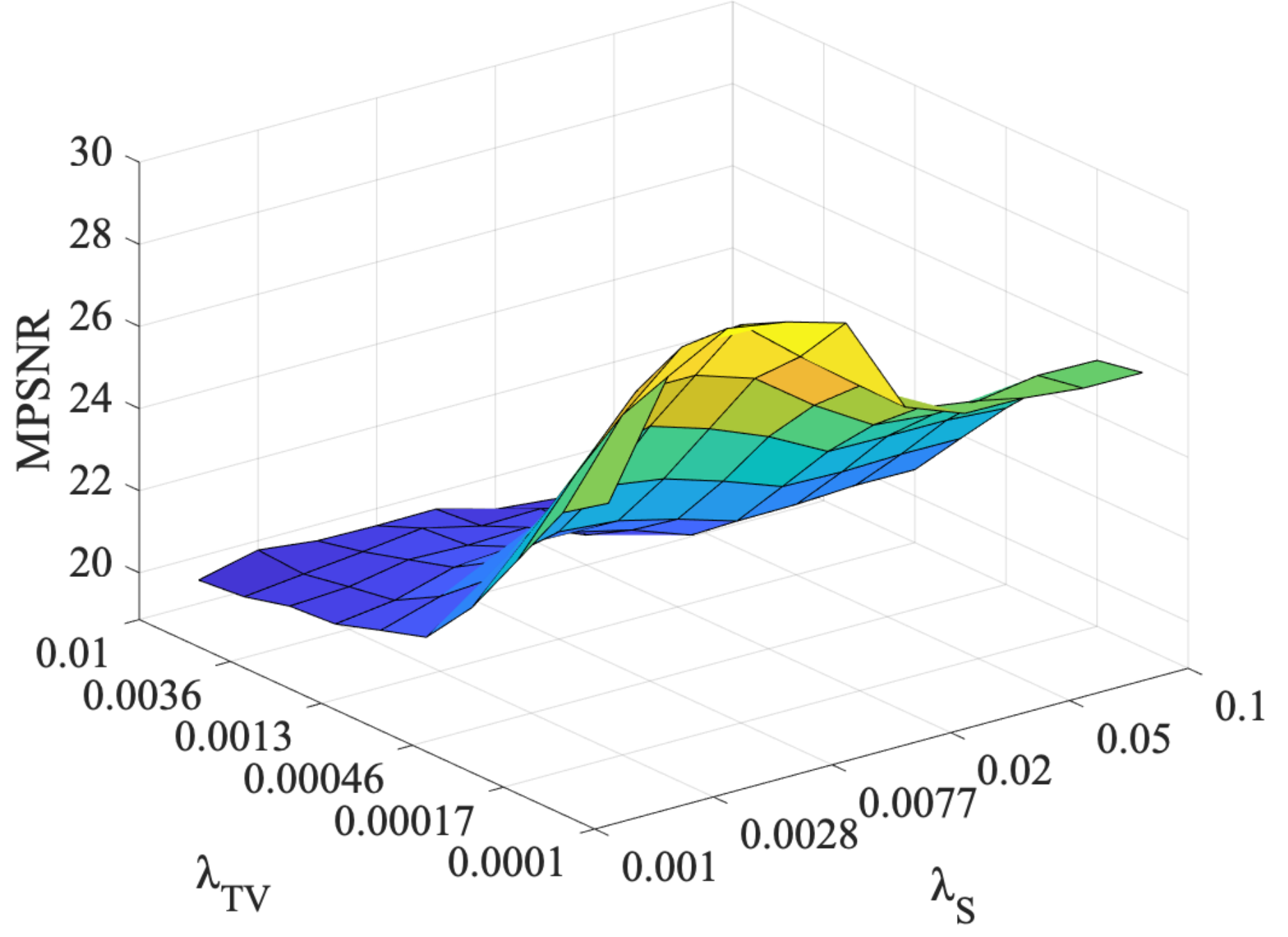}}  
        \subfigure[$\lambda_{\mathcal{N}}=0.018$]{\includegraphics[width=0.18\hsize]{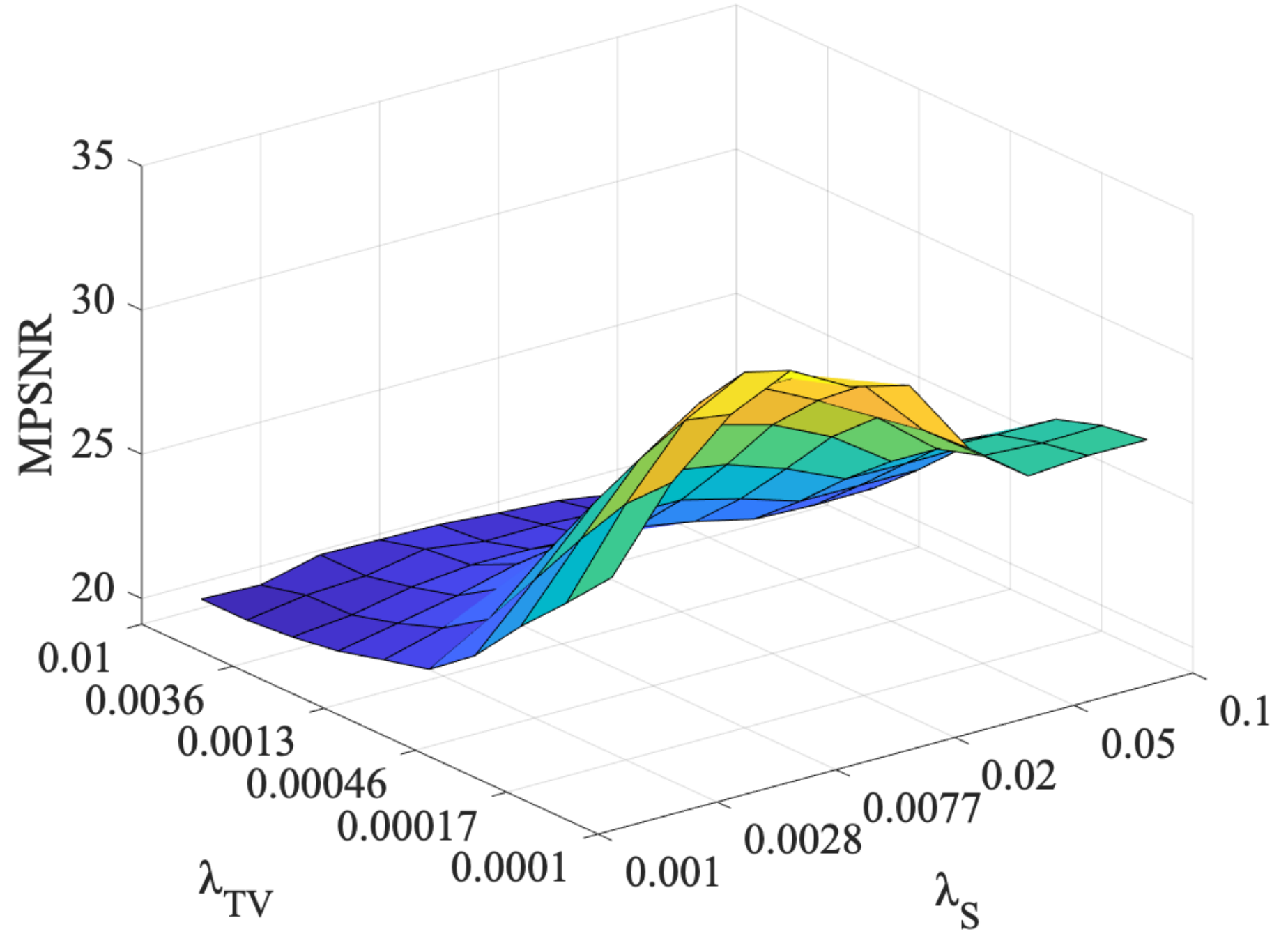}} 
        \subfigure[$\lambda_{\mathcal{N}}=0.032$]{\includegraphics[width=0.18\hsize]{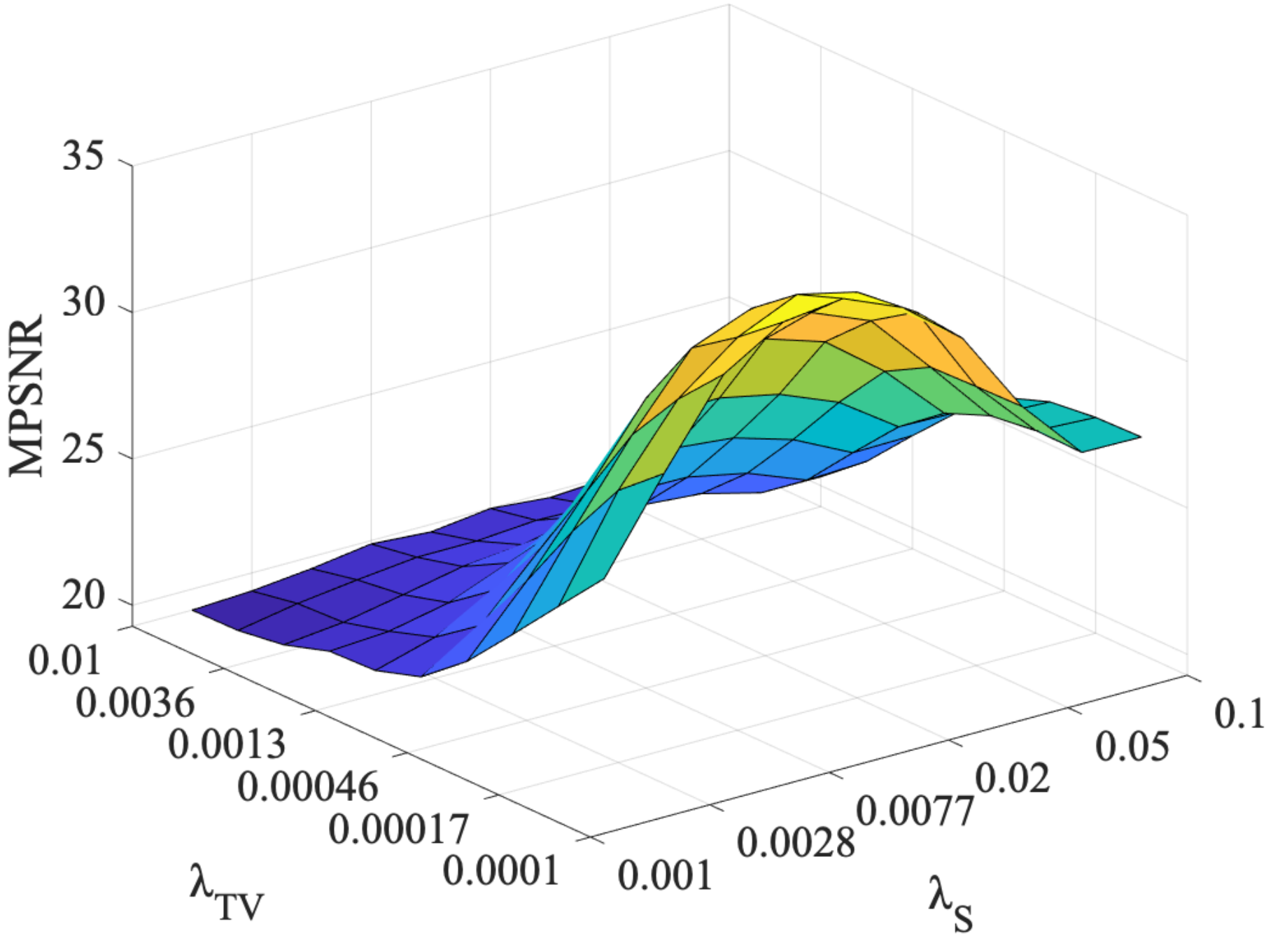}}           
        \subfigure[$\lambda_{\mathcal{N}}=0.056$]{\includegraphics[width=0.18\hsize]{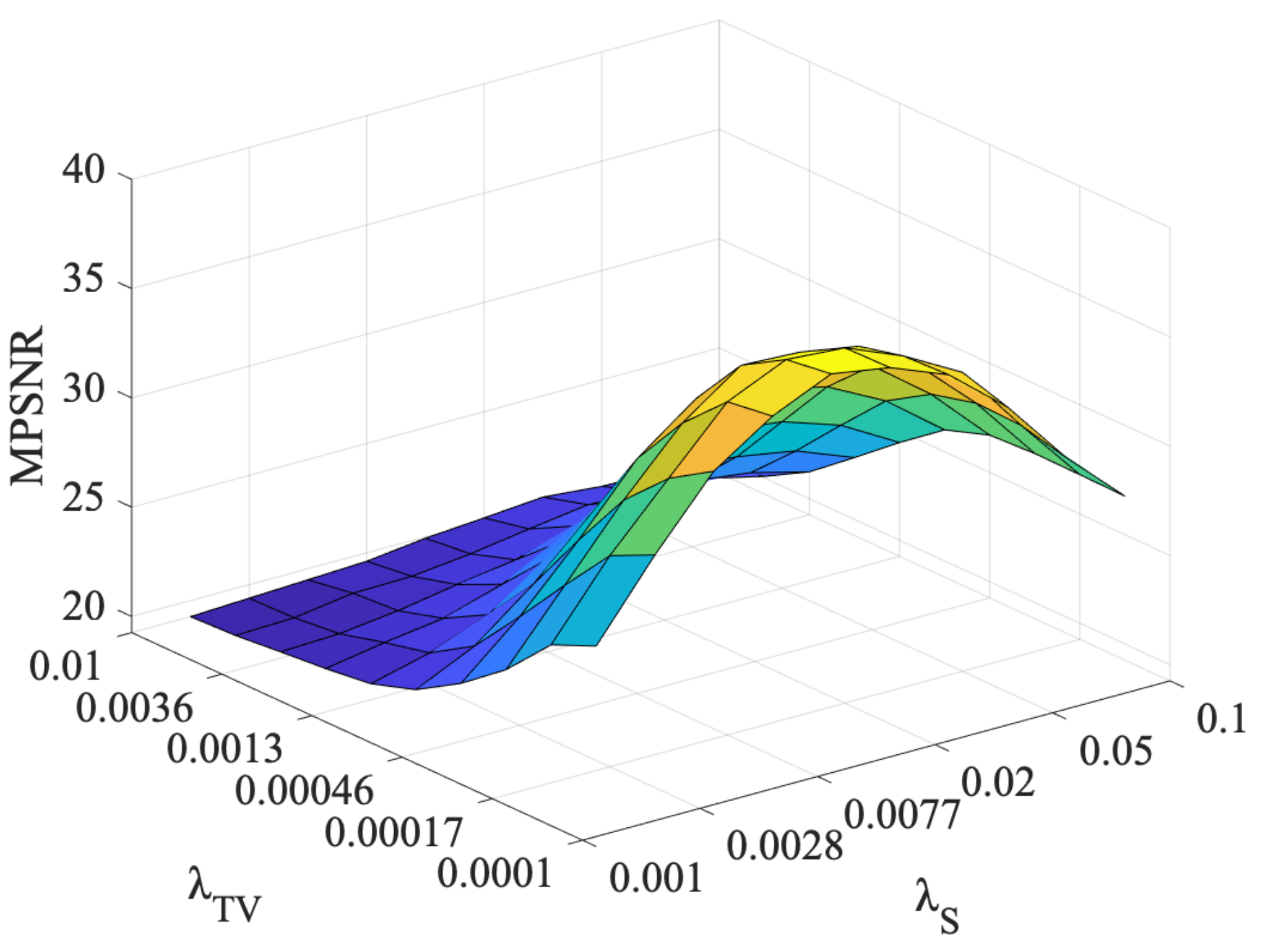}}      
         \subfigure[$\lambda_{\mathcal{N}}=0.1$]{\includegraphics[width=0.18\hsize]{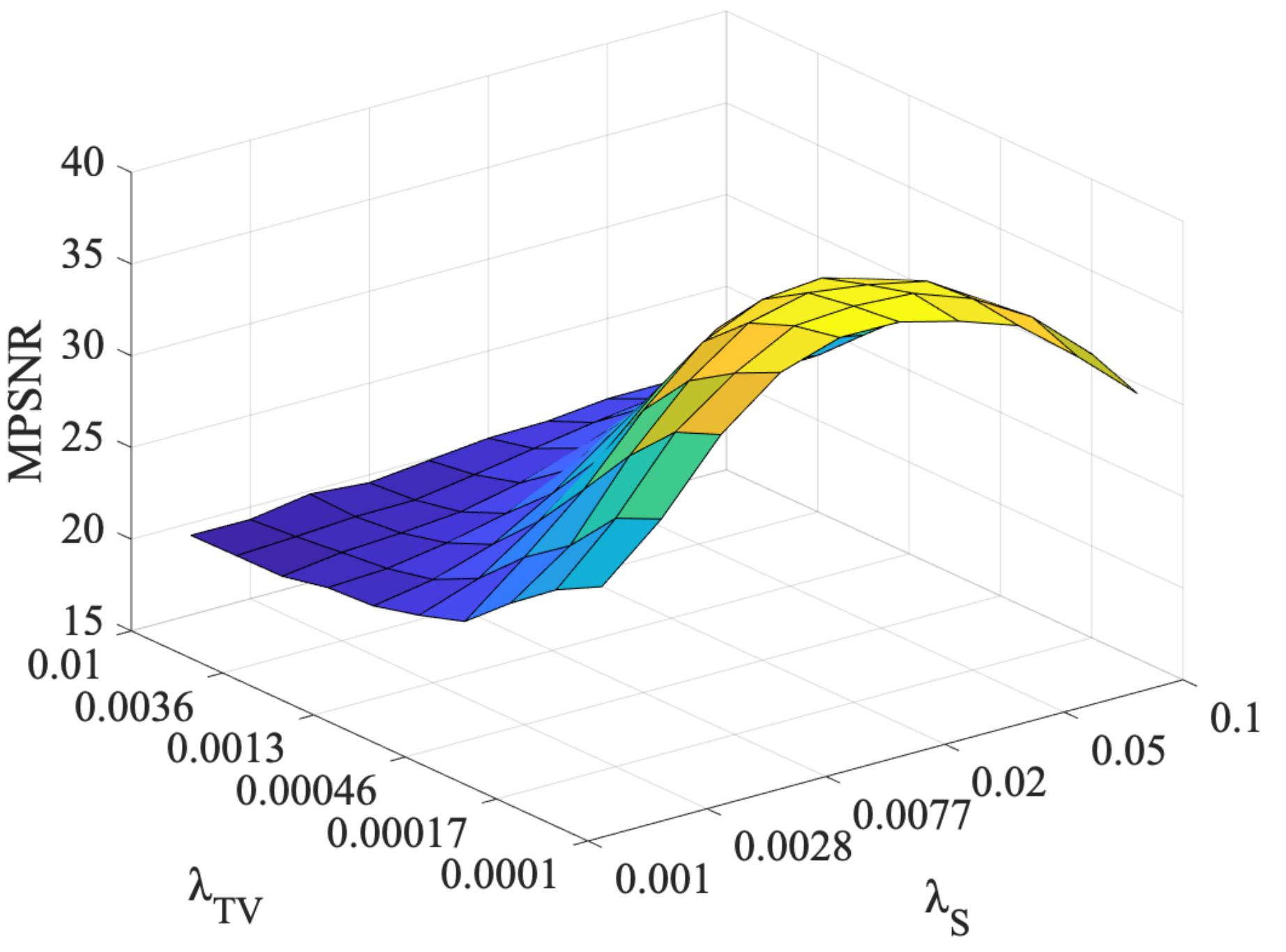}} 
           \caption{The impact of parameters $\lambda_{\text{TV}}$, $\lambda_{\mathcal{S}}$ and $\lambda_{\mathcal{N}}$.}
    \label{fig:parameters}
\end{figure*}

2. The impact of parameters $R$:
\begin{figure}[htbp]
\begin{center}
\includegraphics[width=0.8\hsize]{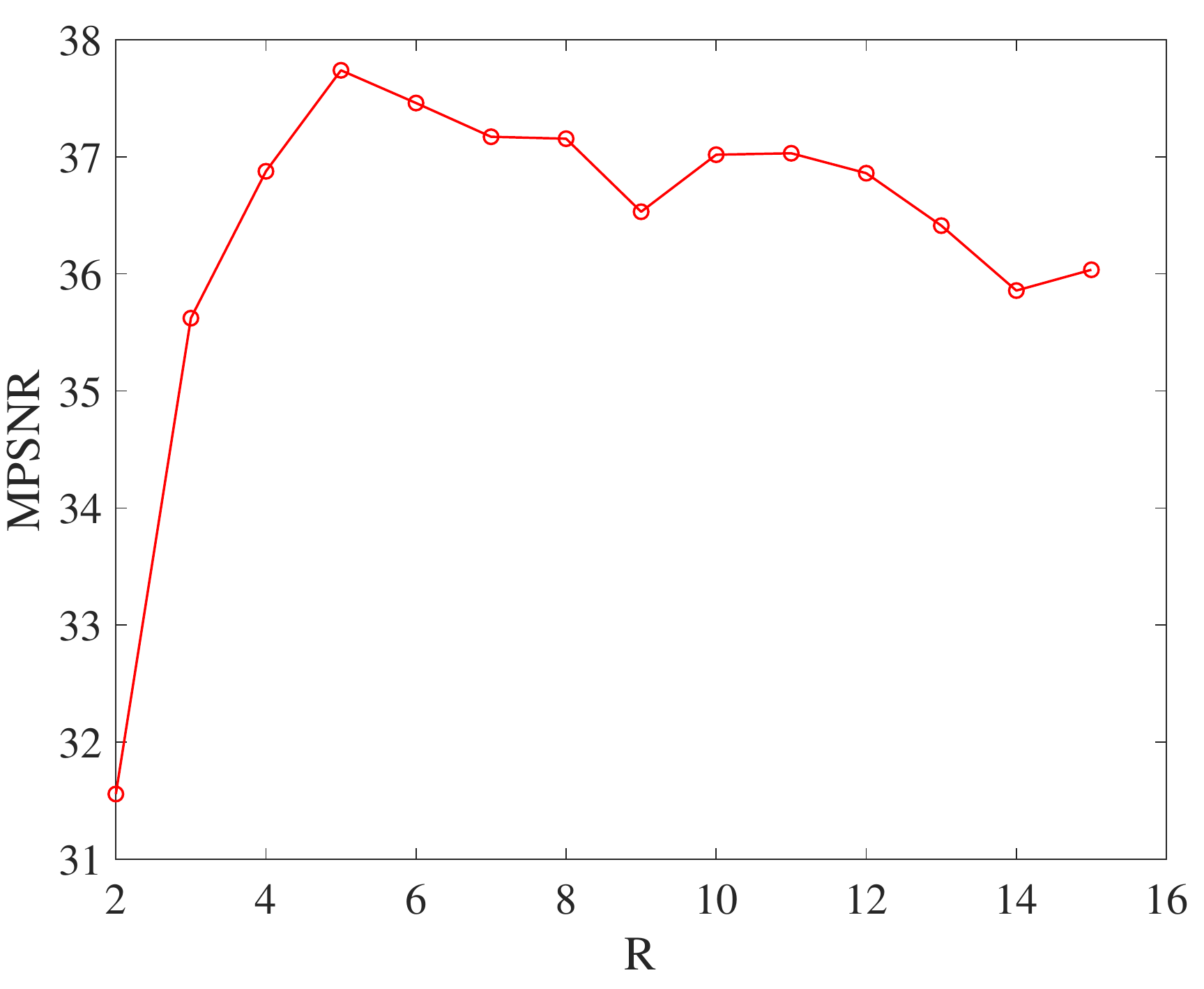}
\caption{The impact of parameters $R$}
\label{fig:R}
\end{center}
\end{figure}
The parameter $R$ in our algorithm represents the number of endmembers.  Fig. \ref{fig:R} shows the MPSNR value as a function of $R$, where $R$ changes from 2 to 15 with step size 1. From it, we can see when $R=5$, the value of MPSNR arrives at its peak.
\section{Conclusion}

In this paper, we develop a new low rank partially orthogonal MVTF model for the hyperspectral image denoising. In addition to the low rank tensor term for global data structure, a 3D tensor total variation is used to exploit the local data structure. The mixed noise is removed by different kinds of norm constraints in the optimization model too. The ADMM is used to efficiently solve the problem. Numerical experiments on HSI denoising show that our algorithm is superior to state-of-the-art algorithms in terms of recovery quality.

\bibliographystyle{ieeetr}
\bibliography{ref}
\end{document}